\documentclass{jfm}

\usepackage{epstopdf, epsfig}
\graphicspath{{./pic/}}
\usepackage{color}
\usepackage{amsmath}
\usepackage{bm}% bold math
\usepackage{tabularx}
\usepackage{float}
\usepackage[singlelinecheck=false, width=\linewidth, justification=justified]{caption}
\usepackage{graphicx}
\usepackage{newtxtext}
\usepackage{newtxmath}
\usepackage{natbib}
\usepackage{hyperref}
\hypersetup{
    colorlinks = true,
    urlcolor   = blue,
    citecolor  = black,
}

\newcommand{\RomanNumeralCaps}[1]
\linenumbers

\shorttitle{Slip-enhanced RP instability of a liquid film on a fibre}
\shortauthor{C. Zhao, Y. Zhang and T. Si}

\title{Slip-enhanced Rayleigh-Plateau instability of a liquid film on a fibre}

\author{Chengxi Zhao\aff{1},
  Yixin Zhang\aff{2}
 \and Ting Si\aff{1} 
 \corresp{\email{tsi@ustc.edu.cn}}}

\affiliation{\aff{1} Department of Modern Mechanics, University of Science and Technology of China, Hefei 230026, China
\aff{2} Physics of Fluids Group and Max Planck Center Twente for Complex Fluid Dynamics, MESA+ Institute and J. M. Burgers Centre for Fluid Dynamics, University of Twente, P.O. Box 217, 7500 AE Enschede,The Netherlands}

\begin{document}
\maketitle

\begin{abstract}
Boundary conditions at a liquid-solid interface are crucial to dynamics of a liquid film coated on a fibre. 
Here a theoretical framework based on axisymmetric Stokes equations is developed to explore the influence of liquid-solid slip on the Rayleigh-Plateau instability of a cylindrical film on a fibre. 
The new model not only shows that the slip-enhanced growth rate of perturbations is overestimated by the classical lubrication model, but also indicates a slip-dependent dominant wavelength, instead of a constant value obtained by the lubrication method, which leads to larger drops formed on a more slippery fibre.   
The theoretical findings are validated by direct numerical simulations of Navier-Stokes equations via a volume-of-fluid method.
Additionally, the slip-dependent dominant wavelengths predicted by our model agree with the experimental results provided by Haefner \textit{et al.}[\textit{Nat. Commun.}, Vol. 6(1), 2015, pp. 1-6].
\end{abstract}


%{\bf MSC Codes }  {\it(Optional)} Please enter your MSC Codes here

%% introduction~~~~~~~~~~~~~~~~~~~~~~~~~~~~~~~~~~~~~~~~~~~~~~~~~~~~~~~~~~~~~~~~~~~~~~~~~~~~~~~~~
\section{Introduction}
Surface-tension-driven instability of liquid jets and their subsequent disintegration into droplets has been much investigated since the pioneering work of \cite{plateau1873} and \cite{rayleigh1878instability,rayleigh1892xvi}. 
This instability also plays an important role in the dynamics of a liquid film coated on a fibre with additional complexities introduced by liquid-solid interfaces, which has received considerable scientific interest \citep{quere1999fluid} because it is crucial to  a range of technologies such as optical fibre manufacturing \citep{deng2011exploration}, droplet transport \citep{lee2022multiple} and water collection through fog harvesting \citep{chen2018ultrafast, zhang2022combinational}.

%  
% review the paper on the instability (logic)
The Rayleigh-Plateau (RP) instability analysis for annular films was first performed by \cite{goren1962instability}, who showed that a film of outer radius $h_0$ is unstable to sufficiently long wavelength disturbances $\lambda > \lambda_\mathrm{crit} = 2 \pi h_0$ ($\lambda_\mathrm{crit}$ is the critical wavelength of the instability beyond which no growth of the instability occurs) and dominant (the most unstable/fast growing) modes depend on the ratio of $h_0$ to the fibre radius $a$, confirmed by experiments \citep{goren1964shape}.
\cite{hammond1983nonlinear} found the same instability of a thin film inside a capillary tube with a nonlinear analysis for the appearance and growth of periodically spaced lobes. 
Under the lubrication approximation that the film thickness
is much smaller than the fibre radius (i.e. $ h_0-a \ll a$), \cite{frenkel1992nonlinear} proposed a weakly nonlinear thin-film equation.
Because of its simplicity and capability of predicting nonlinear behaviours, this lubrication equation and its higher-order versions \citep{craster2006viscous,ruyer2008modelling} have been widely used to study thin-film dynamics flowing down a fibre theoretically \citep{kalliadasis1994drop, yu2013velocity} and explain experimental findings \citep{quere1990thin,duprat2007absolute,craster2009dynamics, ji2019dynamics}.
Recently, these lubrication models have been extended for more complicated situations with other physics taken into account, such as electric fields \citep{ding2014dynamics}, heat transfer \citep{zeng2017experimental}, thermal fluctuations \citep{zhang2021thermal} and Van der Waals forces \citep{tomo2022observation}.

% turn to the slip
Most of the previous works mentioned employed the classical no-slip boundary condition at the liquid-solid interface, which is inaccurate in some practical cases. 
For example, silicone oil on a solid wall made of nylon, used in Qu{\'e}r{\'e}'s experiments \citep{quere1990thin}, has been found to exhibit non-negligible fluid slippage \citep{brochard1994wetting, Lauga2007}.
This slip effect has attracted much research attention recently \citep{secchi2016massive,zhang2020nanoscale,kavokine2021fluids, kavokine2022fluctuation} and been known to be crucial to dynamics of a variety of interfacial flows \citep{liao2014speeding,halpern2015slip,martinez2020effect,zhao2022fluctuation}. 
For a cylindrical film, \cite{ding2011stability} proposed a lubrication equation incorporating slip to explore the instability of a film flowing down a porous vertical fibre. They found that the instability is enhanced by the fluid-porous interface, modelled as a slip boundary condition in the lubrication model.
A two-equation lubrication model was then developed for the case at non-infinitesimal Reynold numbers with inertia taken into account \citep{ding2013viscous}. 
For an annular film inside a slippery tube, \cite{liao2013drastic} solved a lubrication equation with leading order terms numerically and found that the instability is exaggerated by a fractional amount of wall slip, resulting in much more rapid draining compared to the usual no-slip case \citep{hammond1983nonlinear}.
\cite{haefner2015influence} experimentally investigated the influence of slip on the RP instability for a film on a horizontal fibre. Similar to the findings of both \cite{ding2011stability} and \cite{liao2013drastic}, the instability was found to be enhanced by the wall slip with larger growth rates of undulations obtained. Additionally, these experimental results were shown to match predictions of a slip-modified lubrication equation.
\cite{halpern2017slip} later demonstrated that wall slip can enhance the drop formation in a film flowing down a vertical fibre, providing a plausible interpretation for the discrepancy between the experimentally-obtained and theoretically-predicted critical Bond numbers for drop formation.
More recently, the slippage hydrodynamics of cylindrical films were further explored in a non-isothermal environment \citep{chao2018dynamics}
and under the influence of intermolecular forces \citep{ji2019dynamics} via more complicated lubrication models.

% setbacks...
Noticeably, \cite{kliakhandler2001viscous} experimentally demonstrated that the lubrication models for no-slip cases are only valid when the film thickness is smaller than the fibre radius, confirmed by following theoretical analysis performed by \cite{craster2006viscous} based on the Stokes equations.
However, when slip is considered, the accuracy of lubrication models for films on a fibre has never been examined carefully, despite its wide usages in theoretical studies \citep{ding2011stability,liao2013drastic,halpern2017slip,chao2018dynamics} and experimental investigations \citep{haefner2015influence,ji2019dynamics}. In fact, slip-modified  lubrication models predict a dominant wavelength independent of slip length, which somehow is not consistent with experimental data \citep{haefner2015influence}. Thus, it is necessary to provide a more general theoretical analysis to go beyond the lubrication paradigm and clarify the validities of slip-modified lubrication models. For example, it is unclear whether slip-modified lubrication models work for large-slip cases.

% paper structure
In this work, linear stability analysis of the axisymmetric Stokes equations is performed to investigate slip-enhanced RP instability of liquid films on a fibre. Direct numerical simulations of the Navier-Stokes (NS) equations are employed to validate the theoretical findings and provide more physical insights. 
We also compare our theory with available experimental results \citep{haefner2015influence}.
The article is laid out as follows. 
Non-dimensionalised governing equations for films on fibres are introduced in \S\,\ref{sec_MM}. 
In \S\,\ref{sec_instability}, a theoretical model incorporating slip is developed for the RP instability. 
Numerical simulations, performed in \S\,\ref{sec_num}, are compared with the predictions of the theoretical model for the influence of slip on the growth rate (\S\,\ref{subsec_grow_rate}) and the dominant wavelength (\S\,\ref{subsec_dominant}) of perturbations.
In \S\ref{sec_exp}, the theoretically-predicted dominant wavelengths are compared with the experimental data of \cite{haefner2015influence}.

% mathematical model ~~~~~~~~~~~~~~~~~~~~~~~~~~~~~~~~~~~~~~~~~~~~~~~~~~~~~~~~~~~~~~~~~~~~~~~~~~~ 
\section{Model formulation \label{sec_MM}}
\begin{figure}
\centering
\captionsetup{justification=centering}
\includegraphics[width=0.8\textwidth]{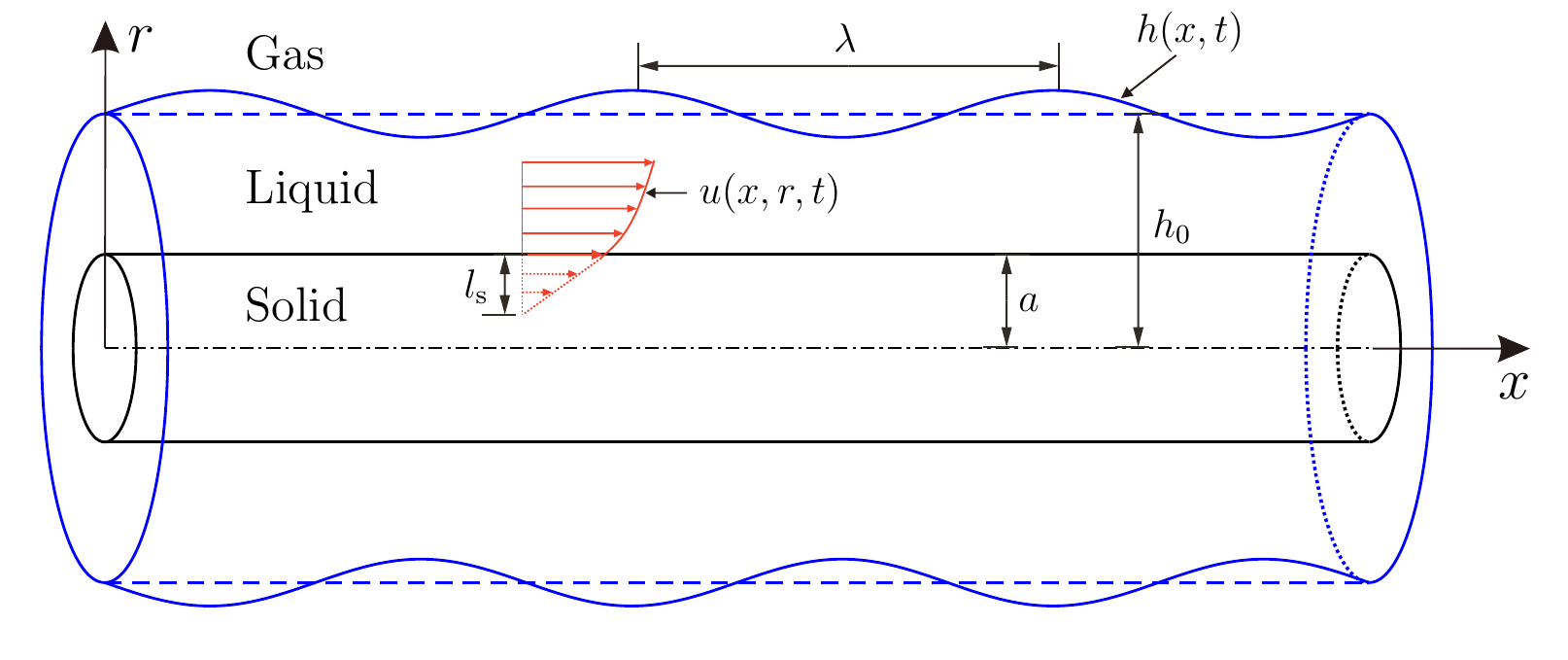}
	\caption{Schematic of a liquid film on a fibre}
\label{fig_schematic}	
\end{figure}

We consider a Newtonian liquid film coating on a horizontal fibre of the radius $a$ with the $x$-axis along the centre line (figure\,\ref{fig_schematic}). 
The initial radius of the film measured from the $x$-axis is $r = h_0$.

The incompressible NS equations are employed to predict the dynamics of the flow inside the liquid film.
To identify the governing dimensionless parameters, we non-dimensionalise the NS equations with the rescaling variables shown below:
\begin{equation}
\label{eq_scaling}
(x,r,h)= \frac{(\tilde{x},\tilde{r},\tilde{h})}{h_0}, 
\quad t  =\frac{\gamma}{\mu h_0} \tilde{t}, 
\quad \mathbf{u} =\frac{\mu}{\gamma} \tilde{\mathbf{u}},
\quad (p,\bm{\tau}) = \frac{h_0}{\gamma}(\tilde{p},\tilde{\bm{\tau}}),
\end{equation}
where $\tilde{h},\,\, \tilde{t},\,\, \tilde{\mathbf{u}},\,\,\textrm{and}\,\, \tilde{p} $ represent the dimensional interface height, time, velocity, and pressure respectively (note that the dimensional material parameters are not given tildes). $\tilde{\bm{\tau}}$ is the shear stress, which is proportional to the strain rate in Newtonian fluids.
$\mu$ is the liquid dynamic viscosity and $\gamma$ is surface tension of the liquid-gas interface.
The dimensionless NS equations can be written as:
\begin{align}
\label{eq_NS_mass} & \nabla \cdot \mathbf{u} =0 \,, \\
 \label{eq_NS_mom} &  \partial_{t} \mathbf{u}+ \mathbf{u} \cdot
\nabla \mathbf{u} = \mathrm{Oh}^2 \left( \nabla \cdot \bm{\tau} - \nabla p \right) \,.
\end{align} 
The non-dimensional quantity $\mathrm{Oh}=\mu/\sqrt{\rho \gamma h_0}$ is the Ohnesorge number, which relates the viscous forces to inertial and surface-tension forces.
Note that, in the previous experiment for the slip-enhanced RP instability \citep{haefner2015influence}, a high viscosity polymer liquid, entangled polystyrene, is coated on a fibre with radius $10 \textrm{ - } 25\,\mu\mathrm{m}$. 
The density of the entangled polystyrene $\rho =1.05\,\mathrm{g\,cm^{-3}}$ and its liquid-gas surface tension $ \gamma = 30.8\,\mathrm{mN\,m^{-1}}$.
The liquid viscosity depends on temperature with the various values in the range $0.2 \textrm{ - } 10\, \mathrm{kg\, m^{-1} s^{-1}} $ \citep{haefner2015rayleigh}. So the Ohnesorge number of liquid film in the experiment is larger than 10 and $\mathrm{Oh}^2 \gg 1$ in equation\,(\ref{eq_NS_mom}). 
It is reasonable to assume that the inertial terms are negligible compared to the viscous terms. 
With axisymmetric initial perturbations, equation\,(\ref{eq_NS_mass}) and (\ref{eq_NS_mom}) can be reduced to axisymmetric Stokes equations, written as:
\begin{equation}
\label{eq_ANS_stokes1}
 \frac{\partial u}{\partial x} + \frac{1}{r}  \frac{\partial (vr)}{\partial r}  = 0 \,,
 \end{equation}
\begin{equation}
 	\label{eq_ANS_stokes2}
 \frac{\partial p}{\partial x} =
 \frac{\partial^2 u}{\partial x^2} + \frac{1}{r}\frac{\partial}{\partial r} \left( r \frac{\partial u}{\partial r} \right) \,, 
\end{equation}
\begin{equation}
  	\label{eq_ANS_stokes3}
   \frac{\partial p}{\partial r} = 
    \frac{\partial^2 v}{\partial x^2} + \frac{\partial}{\partial r} \left[\frac{1}{r} \frac{\partial (vr)}{\partial r}\right]  \,.
\end{equation}
where $u$ and $v$ represent the axial and radial velocity respectively. 

Since the density of gas around the film is much smaller than that of liquid, the gas flow outside can be assumed to be dynamically passive to simplify the problem.  
The liquid-gas interface height $h(x,t)$ satisfies the kinematic boundary condition 
\begin{equation}
\label{eq_stokes4}
\frac{\partial h}{\partial t} + u \frac{\partial h}{\partial x} = v\,.
\end{equation}
The normal stress balances at the interface $r=h$ gives
\begin{equation}
\label{eq_NS_normal}  p- \bm{n} \cdot \bm{\tau} \cdot
\bm{n} = \nabla \cdot \bm{n} \,,
\end{equation}
where $\bm{n}$ is the outward normal and $\nabla \cdot \bm{n}$ represents the dimensionless Laplace pressure.
The tangential force balance is
\begin{equation}
\label{eq_NS_tangential}  \bm{n} \cdot \bm{\tau}  \cdot
\bm{t}=0 \,,
\end{equation}
where $\bm{t}$ is the tangential vector.
With $\bm{n}$ and $\bm{t} $ expressed in terms of the unit vectors in $x$-direction ($\hat{\bm{e}}_x$) and $r$-direction ($\hat{\bm{e}}_r$), 
$$\bm{n} = -\frac{\partial_x h}{\sqrt{1+(\partial_x h)^2}}  \hat{\bm{e}}_x + \frac{1}{\sqrt{1+(\partial_x h)^2}} \hat{\bm{e}}_r \textrm{\,\,\,and\,\,\,} \bm{t} = \frac{1}{\sqrt{1+(\partial_x h)^2}}  \hat{\bm{e}}_x + \frac{\partial_x h}{\sqrt{1+(\partial_x h)^2}}  \hat{\bm{e}}_r,$$
(\ref{eq_NS_normal}) and  (\ref{eq_NS_tangential}) explicitly give
\begin{small} 
  \begin{equation}
  \label{eq_stokes5}
  \begin{split}
  p - \frac{2}{1+ \left(\partial_x h \right)^2} 
  \left[  \frac{\partial v}{\partial r} - \frac{\partial h}{\partial x} \left( \frac{\partial u}{\partial r} 
  + \frac{\partial v}{\partial x}\right) + \left(\frac{\partial h}{\partial x} \right)^2 \frac{\partial u}{\partial x} \right]  
 =\frac{1}{h \sqrt{1+\left(\partial_x h \right)^2}}  - \frac{\partial_x^2 h}{\left(1+\left(\partial_x h \right)^2\right) ^{3/2}}   \,
  \end{split}
  \end{equation} 
  \end{small}
for the normal forces, and
  \begin{equation}
  \label{eq_stokes6}
  2 \frac{\partial h}{\partial x} \left( \frac{\partial v}{\partial r} - \frac{\partial u}{\partial x} \right) 
  + \left[ 1- \left(\frac{\partial h}{\partial x} \right)^2 \right]  \left( \frac{\partial u}{\partial r} + \frac{\partial v}{\partial x} \right) = 0 
  \end{equation}
for the tangential forces. Here $\partial_x$ and $\partial_x^2$ refers to first and second spatial derivatives.

In terms of the boundary conditions at the fibre surface $r = \alpha$ ($\alpha$ is the dimensionless fibre radius, i.e. $\alpha = a/h_0$), we have the slip boundary condition in tangential direction and the no-penetration boundary condition in the normal direction such that
\begin{equation}
\label{eq_stokes7}
u = l_\mathrm{s} \frac{\partial u}{\partial r}, \quad v=0\,.
\end{equation}
Here, $l_\mathrm{s}$ represents the dimensionless slip length rescaled by $h_0$.

The governing equations above can be further simplified under the long-wave approximation to give a one-dimensional lubrication equation, compared with experimental results in the previous work \citep{haefner2015influence}.
This lubrication equation is displayed here for the instability analysis appearing in \S\,\ref{sec_instability} (see Appendix\,\ref{app_LE_deri} for the derivation).
Its dimensionless format is
\begin{equation}
\label{eq_LE_frame}
\frac{\partial h}{\partial t} = \frac{1}{ h} \frac{\partial}{\partial x} \left[ M(h) \frac{\partial p}{\partial x} \right]\,,
\end{equation}
where $M(h)$ is the the mobility term
\begin{equation}
M(h) =\frac{1}{16} \left[-3h^4-\alpha^4 + 4 \alpha^2 h^2 + 4 h^4 \mathrm{ln}\left(\frac{h}{\alpha}\right) + 4 l_\mathrm{s} \frac{(h^2-\alpha^2)^2}{\alpha} \right]\,,
\end{equation}
and $p$ is the Laplace pressure
\begin{equation}
\label{eq_LE_laplace_pressure}
p  =  \frac{1}{h}  - \frac{\partial^2 h}{\partial x^2} \,.
\end{equation}

\section{Instability analysis \label{sec_instability}}
In this section, linear instability analysis based on equations\,(\ref{eq_ANS_stokes1} - \ref{eq_stokes7}) is performed using the normal mode method, which has been widely used for the RP instability in different fluid configurations \citep{rayleigh1878instability, tomotika1935instability, craster2006viscous, liang2011linear}.
The dimensionless perturbed quantities are expressed as
$$u(x,r,t) = \hat{u}(r) e^{\omega t+ikx}, 
v(x,r,t) = \hat{v}(r) e^{\omega t+ikx}  \textrm{ and } 
p(x,r,t) = 1 + \hat{p}(r) e^{\omega t+ikx}\,,
$$ 
where $\omega$ is the growth rate of perturbations and $k$ is the wavenumber.
Substituting these quantities into the mass and momentum equations (\ref{eq_ANS_stokes1} - \ref{eq_ANS_stokes3}) leads to
\begin{align}
\label{eq_stokes_linear3}
ik \hat{u} & + \frac{1}{r}  \frac{d (\hat{v} r)}{d r}  = 0 \,, \\
\label{eq_stokes_linear1}
ik \hat{p} & =  -k^2 \hat{u} + \frac{1}{r}\frac{d}{d r} \left( r \frac{d \hat{u}}{d r} \right) \,, \\
\label{eq_stokes_linear2}
\frac{d \hat{p}}{d r} & =  -k^2 \hat{v} + \frac{d}{d r}\left[\frac{1}{r} \frac{d (\hat{v} r)}{d r}\right] \,. 
\end{align}
With equations\,(\ref{eq_stokes_linear3} - \ref{eq_stokes_linear2}), we can eliminate $\hat{u}$ and $\hat{p}$ to get a fourth order ordinary partial differential equation for $\hat{v}$
\begin{equation}
\label{eq_stokes_linear4}
\frac{d }{d r}\left\lbrace\frac{1}{r} \frac{d}{d r} \left[r \frac{d}{d r} \left(\frac{1}{r} \frac{d(\hat{v}r)}{dr} \right) \right] \right\rbrace - 2k^2 \frac{d}{dr}\left[\frac{1}{r} \frac{d(\hat{v}r)}{dr} \right]+k^4\hat{v} = 0 \,.
\end{equation}
The general solution for (\ref{eq_stokes_linear4}) can be obtained in terms of Bessel functions, written as 
\begin{equation}
\label{eq_stokes_solution1}
\hat{v} = C_1 r K_0(kr) + C_2 K_1(kr)+ C_3 r I_0(kr) + C_4 I_1(kr) \,,
\end{equation}
where $I$ and $K$ are modified Bessel function of first and second kind, The subscripts `$0$' and `$1$' represent order of the Bessel functions. 
$C_1 \textrm{ - } C_4$ are four arbitrary constants to be determined by the boundary conditions and  $r\in[\alpha,1]$.
Substituting (\ref{eq_stokes_solution1}) into (\ref{eq_stokes_linear3}) and (\ref{eq_stokes_linear1}) gives us the solutions for $\hat{u}$ and $\hat{p}$
\begin{align}
\label{eq_stokes_solution2}
\hat{u} & = \left[
C_1 \left(  krK_1-2K_0 \right)+ C_2 kK_0 - 
 C_3\left(2 I_0 + kr I_1 \right) - C_4 kI_0
\right] / (ik) \,, \\
\label{eq_stokes_solution3}
\hat{p} & = 2 \left( C_1 K_0 + C_3 I_0 \right) \,.
\end{align}

In a similar approach, the dimensionless perturbed quantities for $\hat{u}, \hat{v}, \hat{p}$, combined with $h(x,t) = 1 + \hat{h}e^{\omega t+ikx}$, are substituted into perturbed boundary equations (\ref{eq_stokes4} - \ref{eq_stokes7}). 
For the boundary conditions at the interface ($r=1$), their linearisation gives 
\begin{align}
\label{eq_stokes_bc3}
& \frac{d \hat{u}}{d r} + ik \hat{v}  = 0, \\
\label{eq_stokes_bc4}
& \hat{p} - 2 \frac{d \hat{v}}{d r}  
 = \hat{h} \left(k^2-1 \right)\,, \\
\label{eq_stokes_bc5}
& \omega \hat{h} = \hat{v} \,.
\end{align}
And for the boundary conditions on the fibre surface ($r=\alpha$), their linearised forms are
\begin{align}
\label{eq_stokes_bc1}
\hat{u} &= l_\mathrm{s} \frac{d \hat{u}}{d r}\,, \\
\label{eq_stokes_bc2}
\hat{v} &= 0 \,.
\end{align}
According to equation\,(\ref{eq_stokes_bc5}), $\hat{h}$ in (\ref{eq_stokes_bc4}) can be eliminated to give the final four equations of the boundary conditions, i.e. (\ref{eq_stokes_bc3}), (\ref{eq_stokes_bc4}), (\ref{eq_stokes_bc1}) and (\ref{eq_stokes_bc2}). 
Substituting the Bessel functions (\ref{eq_stokes_solution1} - \ref{eq_stokes_solution3}) into these perturbed equations leads to a homogeneous system of linear equations for $C_1 \textrm{ - } C_4$, which has a non-trivial solution only if the determinant of the coefficients vanishes.
In this way, we have the final equation
\begin{equation}
\begin{array}{|cccc|}
F_{11} & F_{12}  & F_{13} & F_{14} \\ 
F_{21} & F_{22}  & F_{23} & F_{24} \\ 
\alpha K_0(k \alpha) & K_1(k \alpha) & \alpha I_0(k \alpha) & I_1(k \alpha) \\
k K_0(k) - K_1(k) & k K_1(k) & k I_0(k) + I_1(k) & k I_1(k) \\
\end{array} =0 \,, 
\end{equation}
where
\begin{align}
F_{11} & = (k^2-1) K_0(k) - 2 \omega k K_1(k)\,, \nonumber \\
F_{12} & =  (k^2-1) K_1(k) - 2 \omega \left[ k K_0(k) + K_1(k) \right]\,, \nonumber \\
F_{13} & = (k^2-1) I_0(k) + 2 \omega k I_1(k)  \,, \nonumber \\
F_{14} & = (k^2-1) I_1(k) + 2 \omega \left[ kI_0(k) - I_1(k) \right]\,, \nonumber \\
F_{21} & = (2-l_\mathrm{s} k^2 \alpha) K_0(k \alpha) + k(2 l_\mathrm{s}-\alpha) K_1(k \alpha) \,, \nonumber \\
F_{22} & = -kK_0(k \alpha) - l_\mathrm{s} k^2 K_1(k \alpha)\,, \nonumber \\
F_{23} & = (2-l_\mathrm{s} k^2 \alpha) I_0(k \alpha) -k(2 l_\mathrm{s}-\alpha) I_1(k \alpha) \,, \nonumber \\
F_{24} & = k I_0(k \alpha) - l_\mathrm{s} k^2 I_1(k \alpha)\,. \nonumber 
\end{align}
Because $\omega$ only appears linearly in the first line of the determinant, the dispersion relation between $\omega$ and $k$ can be expressed explicitly
\begin{equation}
\label{eq_dispersion_stokes}
\omega = \frac{k^2-1}{2} 
\frac{K_0(k) \Delta_1 - K_1(k) \Delta_2 + I_0(k) \Delta_3 -I_1(k) \Delta_4}{k K_1(k) \Delta_1- \left[ kK_0(k)+K_1(k) \right] \Delta_2 -k I_1(k) \Delta_3 + 
\left[ k I_0(k) -I_1(k) \right] \Delta_4} \,,
\end{equation} 
where
\begin{align}
\Delta_1& = 
\begin{array}{|ccc|}
 F_{22} & F_{23} & F_{24} \\
 K_1(k \alpha) & \alpha I_0(k \alpha) & I_1(k \alpha) \\
 k K_1(k) & k I_0(k) + I_1(k) & k I_1(k) \\
\end{array}\,, \nonumber \\ \nonumber
\Delta_2 &= 
\begin{array}{|ccc|}
 F_{21} & F_{23} & F_{24} \\
\alpha K_0(k \alpha)  & \alpha I_0(k \alpha) & I_1(k \alpha) \\
k K_0(k) - K_1(k)  & k I_0(k) + I_1(k) & k I_1(k) \\
\end{array}\,, \\ \nonumber
\Delta_3 &= 
\begin{array}{|ccc|}
 F_{21} & F_{22} & F_{24} \\
\alpha K_0(k \alpha) & K_1(k \alpha)  & I_1(k \alpha) \\
k K_0(k) - K_1(k) & k K_1(k) &  k I_1(k) \\
\end{array}\,, \\ \nonumber
\Delta_4 &= 
\begin{array}{|ccc|}
 F_{21} & F_{22} & F_{23} \\
\alpha K_0(k \alpha) & K_1(k \alpha) & \alpha I_0(k \alpha)  \\
k K_0(k) - K_1(k) & k K_1(k) & k I_0(k) + I_1(k)  \\
\end{array}\,.  
\end{align}
\begin{figure}
\centering
\includegraphics[width=1.0\textwidth]{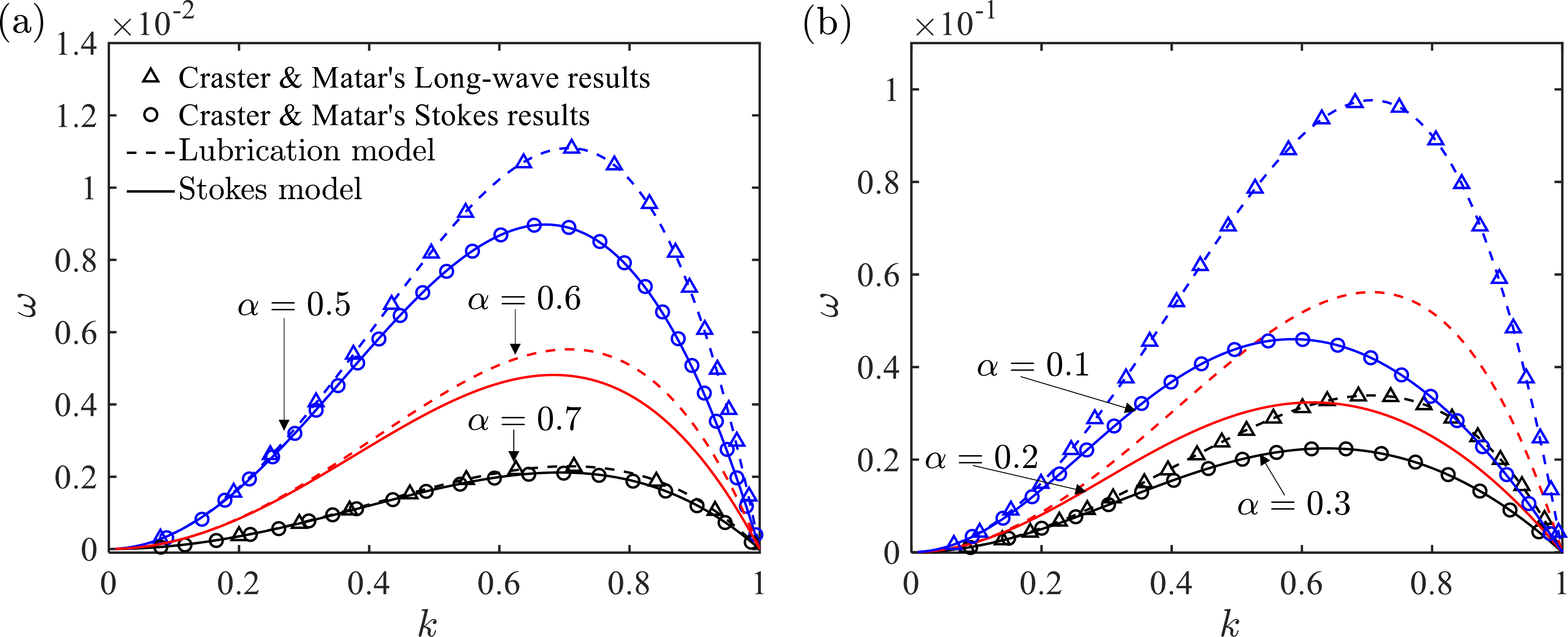}
	\caption{The dispersion relation between the growth rate $\omega$  and the wavenumber $k$ on the no-slip boundary condition ($l_\mathrm{s}=0$) for various dimensionless fibre radii: (a) $\alpha=0.7 \textrm{ (black), } 0.6 \textrm{ (red), } 0.5 \textrm{ (blue)}$; (b) $\alpha=0.3 \textrm{ (black), } 0.2 \textrm{ (red), } 0.1 \textrm{ (blue)}$. The predictions of the Stokes model (\ref{eq_dispersion_stokes}) are plotted in solid lines and results from the lubrication model (\ref{eq_dispersion_LE}) are illustrated in dashed lines. The triangles and circles are the long-wave and Stokes results of \cite{craster2006viscous}, employed here to verify (\ref{eq_dispersion_stokes}) and (\ref{eq_dispersion_LE}), respectively. }
\label{fig_dispersion1}	
\end{figure}

To compare the Stokes model (\ref{eq_dispersion_stokes}) with the lubrication theory, equation\,(\ref{eq_LE_frame}) is linearised using $h = 1 + \hat{h} e^{\omega t + ikx}$, providing the dispersion relation
\begin{equation}
\label{eq_dispersion_LE}
\omega = (k^2-1)  k^2 M,
\end{equation}
where 
$M =\left[3+ \alpha^4 - 4 \alpha^2 + 4\, \mathrm{ln}\,\alpha - 4 l_\mathrm{s} (1-\alpha^2)^2/\alpha \right]/16.$
\begin{figure}
\centering
\includegraphics[width=1.0\textwidth]{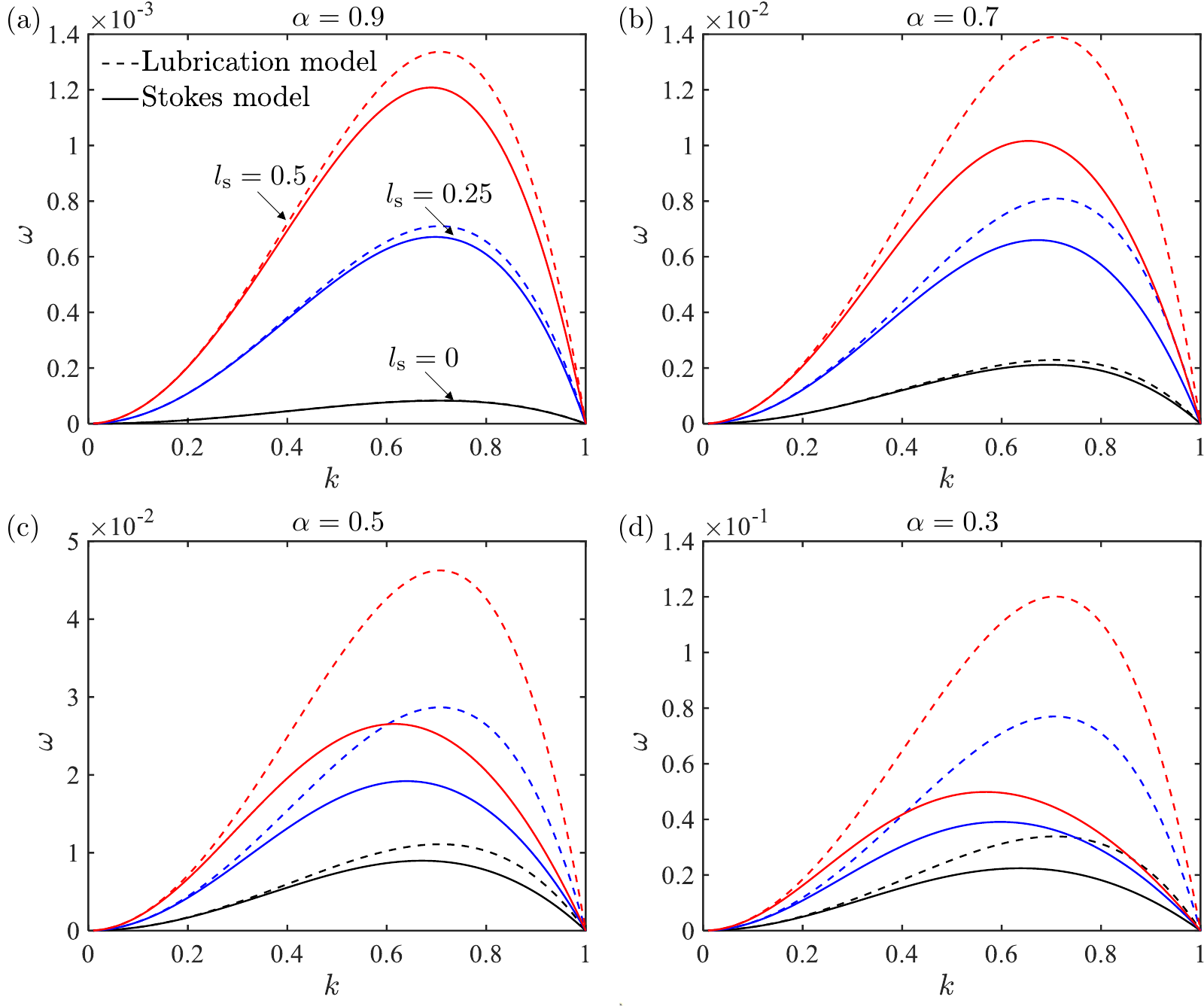}
	\caption{The dispersion relation between the growth rate $\omega$  and the wavenumber $k$ on different boundary conditions of various dimensionless fibre radii: (a) $\alpha=0.9$; (b) $\alpha=0.7$; (c) $\alpha=0.5$; (d) $\alpha=0.3$.  The solid and dashed lines are the predictions of the Stokes model (\ref{eq_dispersion_stokes}) and the lubrication model (\ref{eq_dispersion_LE}), respectively, on three boundary conditions with $l_\mathrm{s}=0 \textrm{ (black), } 0.25 \textrm{ (blue) and } 0.5 \textrm{ (red).}$  }
\label{fig_dispersion2}	
\end{figure}

In figure\,\ref{fig_dispersion1}, predictions of (\ref{eq_dispersion_stokes}) and (\ref{eq_dispersion_LE}) in the limiting case ($l_\mathrm{s}=0$) are found to agree with \cite{craster2006viscous}'s results. 
Note that the lubrication dispersion relation (dashed lines) matches the Stokes dispersion relation (solid lines) when $ \alpha \geq 0.6$.
However, the agreement deteriorates for the smaller $\alpha$, mainly due to the limitations of lubrication approximations for `thicker' films.
Despite the failure in predicting the dispersion relation for small $\alpha$, the Plateau instability criterion with critical wavenumber, $k_\mathrm{crit}=2\,\pi h_0/\lambda_\mathrm{crit}=1$ \citep{plateau1873} can be predicted by the lubrication model for all $\alpha$ values.
The reason is that both the circumferential and tangential curvature for the Laplace pressure is kept in (\ref{eq_LE_laplace_pressure}) to give the term $k^2-1$ in the dispersion relation.

Figure\,\ref{fig_dispersion2} illustrates the dispersion relations of different $l_\mathrm{s}$ values.  
Both the Stokes model and the lubrication model show that the RP instability is enhanced by slip with faster growing perturbations for larger $l_\mathrm{s}$, consistent with previous findings based on lubrication models \citep{liao2013drastic,haefner2015influence, halpern2017slip}.
However, obvious discrepancies between the Stokes dispersion relation and the lubrication dispersion relation are found in most cases in figure\,\ref{fig_dispersion2}.
Even for thin films ($\alpha \geq 0.7$), where the lubrication model has been shown to perform well on the no-slip boundary condition (see the agreement of black lines in figure\,\ref{fig_dispersion2}\,(a,b)), remarkable discrepancies appear when $l_\mathrm{s} \geq 0.5$, indicating that the lubrication model (\ref{eq_LE_frame}) is not suitable for large-slip cases.
The main reason is that the lubrication approximation imposes a restriction on $l_\mathrm{s}$, namely $l_\mathrm{s} \ll \lambda^2 -1+\alpha$, where $\lambda$ is the perturbation wavelength (see the derivation of (\ref{eq_ls_constraint}) and its corresponding explanation in Appendix\,\ref{app_LE_deri}). 
Moreover, the discrepancy increases for larger $\alpha$, similar to the trend found in the no-slip cases in figure\,\ref{fig_dispersion1}.

\begin{figure}
\centering
\includegraphics[width=1.0\textwidth]{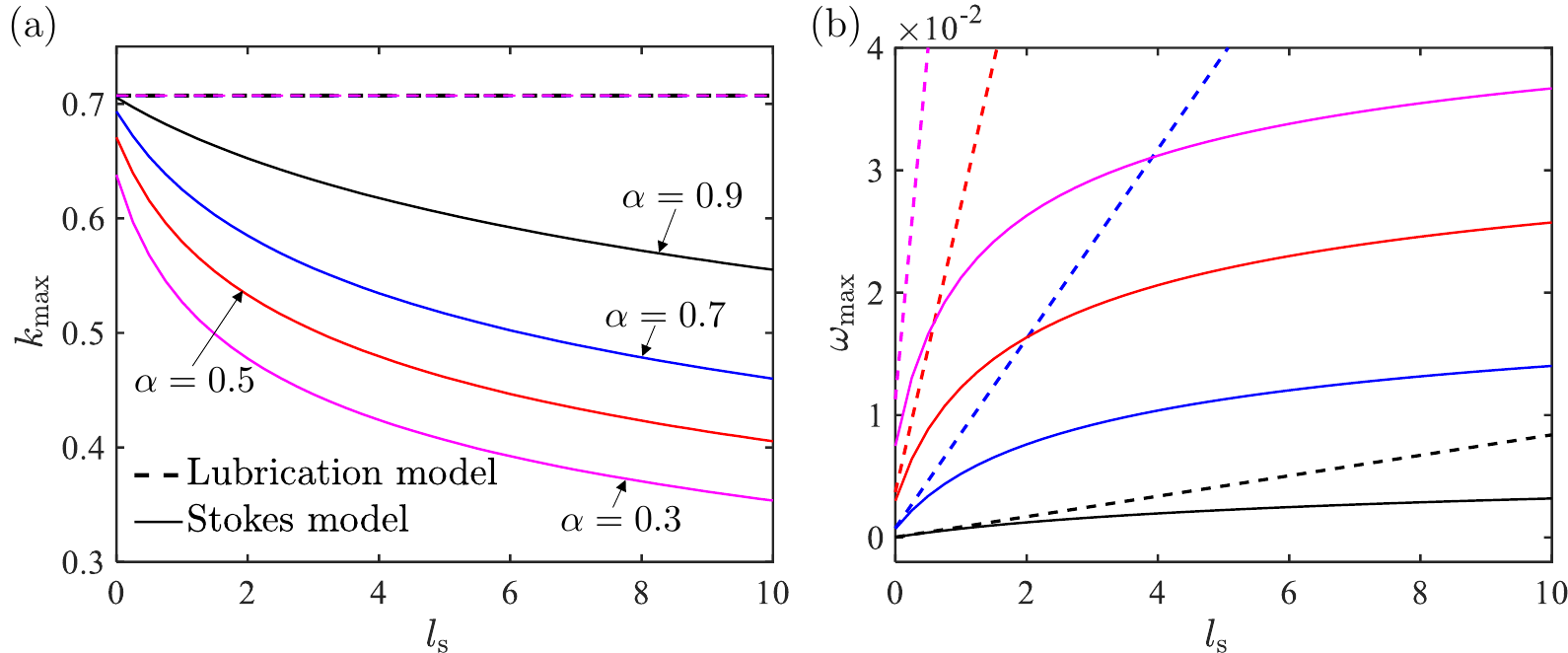}
	\caption{Variations of the dominant wavenumber $k_\mathrm{max}$ and their corresponding growth rate $\omega_\mathrm{max}$ with slip lengths. The solid and dashed lines are the predictions of the Stokes model (\ref{eq_dispersion_stokes}) and the lubrication model (\ref{eq_dispersion_LE}), respectively, for different fibre thicknesses, i.e $\alpha=0.9 \textrm{ (black), } 0.7 \textrm{ (blue), } 0.5 \textrm{ (red) and } 0.3 \textrm{ (purple)}$. }
\label{fig_kmax}	
\end{figure}

The $k_\mathrm{crit}$ values for the RP instability in figure\,\ref{fig_dispersion2} are found not to be affected by the slip.
These values are determined by the competition between the two curvature terms for the Laplace pressure, shown on the right-hand side of (\ref{eq_stokes5}), where the term of the circumferential curvature, $1/ \left(h \sqrt{1+\left(\partial_x h \right)^2} \right)$ is the driving force and the term of tangential curvature, $\partial_x^2 h/ \left(1+\left(\partial_x h \right)^2\right) ^{3/2}$ is the resisting force.
The balance of the two forces gives the  $k_\mathrm{crit}$, which is independent of $l_\mathrm{s}$.
The dominant wavenumber $k_\mathrm{max}$ predicted by the Stokes model in figure\,\ref{fig_dispersion2} is shown to be modified by slip, while $k_\mathrm{max}$ from the lubrication model is unchanged with an analytical expression $k_\mathrm{max} = \sqrt{1/2}$, derived from (\ref{eq_dispersion_LE}).  
The variations of $k_\mathrm{max}$ with slip are further illustrated in figure\,\ref{fig_kmax}\,(a), where $k_\mathrm{max}$ declines with increasing $l_\mathrm{s}$, possibly leading to larger drop sizes after the film breakup.
This trend holds for different fibre radii with more rapid decrease of $k_\mathrm{max}$ of thicker films (smaller $\alpha$).
These findings indicate the limitations of the classical slip lubrication models \citep{liao2013drastic,haefner2015influence,
halpern2017slip,chao2018dynamics,ji2019dynamics}, which can be severe with large slip, while the Stokes model proposed here has the capability of predicting the instability over a much larger range of $l_\mathrm{s}$.
Additionally, the lubrication model shows that corresponding growth rates $\omega_\mathrm{max}$ of each $k_\mathrm{max}$ grow linearly with increasing $l_\mathrm{s}$, which can be described by $\omega_\mathrm{max} = -M/4$,
while the Stokes model presents nonlinear growth with smaller $\omega_\mathrm{max}$. 

\section{Numerical simulations \label{sec_num}}
In this section, direct numerical simulations are performed to support the theoretical findings in \S\,\ref{sec_instability} and provide more physical insights into the slip-enhanced RP instability of the cylindrical films on fibres.
The open-source framework Basilisk \citep{popinet2014basilisk,popinet2018numerical} is employed to solve axisymmetric incompressible Navier-Stokes equations with surface tension.
The interface between  the high-density liquid and the low-density ambient air is reconstructed by a volume-of-fluid method, which has been validated by different kinds of interfacial flows such as jet breakup \citep{deblais2018viscous}, bubble bursting \citep{berny2020role} and 
breaking waves \citep{mostert2020inertial}.

The simulation domain is a rectangle (the section of a hollow cylinder in cylindrical coordinates) with a size $[\alpha, 2]\times [0, L]$. Here $L$ is the length of the film/fibre and $\alpha$ is the radius of the fibre.
The liquid film of initial radius $h_0 = 1$ is placed at the bottom of the domain with small perturbations at the liquid-gas interface.
The left and right boundary conditions of the simulation domain are considered periodic. The top is a symmetry boundary and bottom is a slip-wall boundary, as described by equation (\ref{eq_stokes7}).
All simulations are performed in the dimensionless unit with the rescaling variables displayed in equation\,(\ref{eq_scaling}).
The non-dimensional parameter $\mathrm{Oh}=10$, which comes from liquid properties displayed in the previous experiment \citep{haefner2015rayleigh} with $h_0 = 10\,\mu\mathrm{m},\,\rho =1.05\,\mathrm{g\,cm^{-3}}, \, \gamma = 30.8\,\mathrm{mN\,m^{-1}}\,\,\mathrm{and}\,\, \mu=0.22\,\mathrm{kg\, m^{-1} s^{-1}}$. 
Two configurations with different film lengths and initial perturbations are simulated here to investigate the influence of slip on the growth rates of perturbations (\S\,\ref{subsec_grow_rate}) and the dominant wavelength (\S\,\ref{subsec_dominant}) respectively.

\subsection{Influence of slip on perturbation growth rates \label{subsec_grow_rate}}

\begin{figure}
\centering
\includegraphics[width=1.0\textwidth]{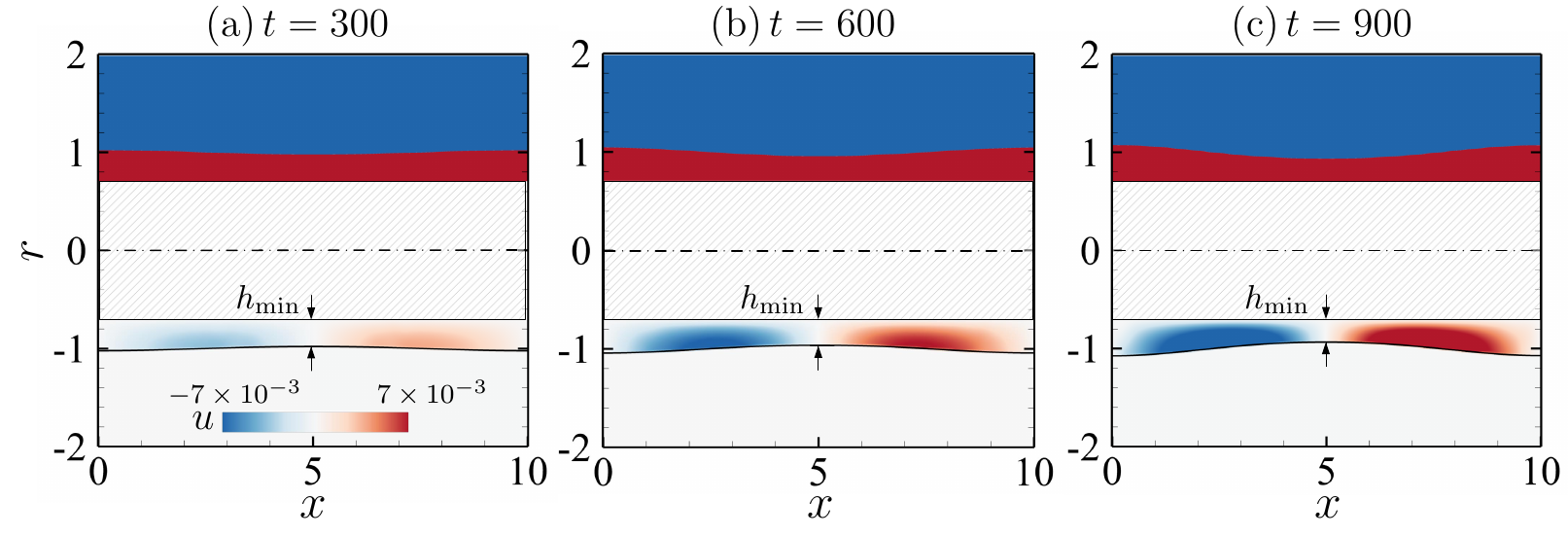}
	\caption{Film thinning simulation results of one perturbation wave in an axially symmetric domain. 
For this case, $L=10$, $l_\mathrm{s}=0$ and $\alpha=0.7$.
	Interface profiles $h$ (upper half) and contours of the axial velocity  $u(x,r,t)$ inside the film (lower half) are shown at three time instants: (a) $t_1=300$; (b) $t_2=600$; (c) $t_3=900$. $h_\mathrm{min}$ represents the minimum radius of the film.
	}
\label{fig_Basilisk_setting}	
\end{figure}

To study the influence of slip on the growth rates, we consider simulations of a relatively short film $L=10$ with an initial perturbation $h(x,t) =1+ \varepsilon\, \mathrm{cos}\left[2 \pi (x/L-1/2)\right]$, where $\varepsilon=0.01$.
High-density grids are employed to capture the interface position and velocity profiles inside the film with $2^{10}$ grid points along the $x$-axis. 
We choose three slip lengths, i.e. $l_\mathrm{s}=0$, $0.25$ and $0.5$ for two fibre radii $\alpha=0.7$ and $\alpha=0.5$. 

\begin{figure}
\includegraphics[width=1.01\textwidth]{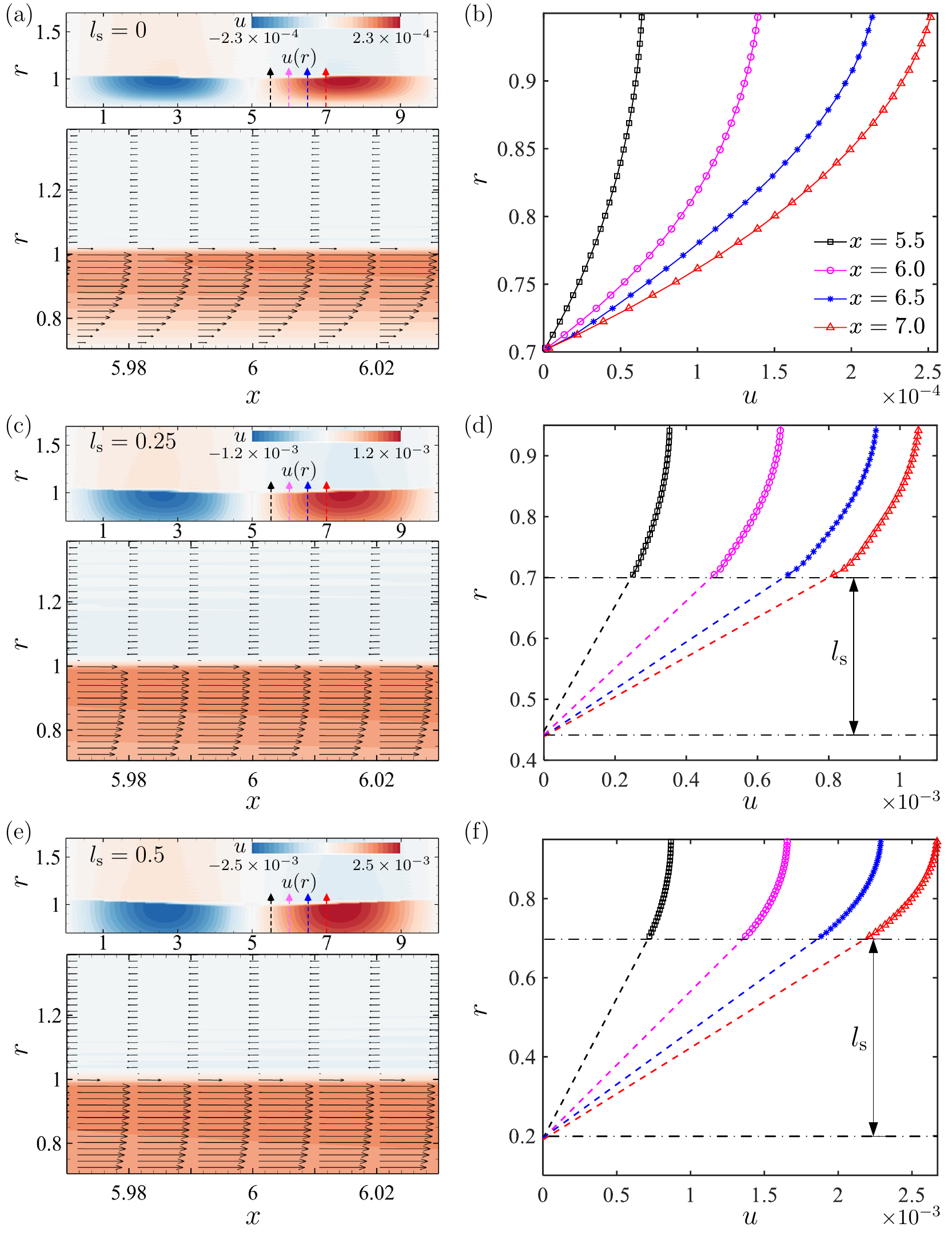}
	\caption{Axial velocity fields at $t=100$ for different slip lengths: (a,b) $l_\mathrm{s} = 0$; (c,d) $l_\mathrm{s} = 0.25$; (e,f) $l_\mathrm{s} = 0.5$. Here, the fibre radius $\alpha = 0.7$.
	The left panel (a,c,e) illustrates the contours of the entire configuration (upper half) and velocity vectors of the local field near $x=6$ (lower half). 
	The right panel shows velocity profiles at four local positions: $x = 5.5$ (black), 6.0 (purple), 6.5 (blue) and 7.0 (red), shown by dashed lines with arrows in the left panel.   
	}
\label{fig_VF}	
\end{figure}

Figure\,\ref{fig_Basilisk_setting} shows the simulation results on a no-slip fibre of the radius $\alpha=0.7$.
The upper half of figure\,\ref{fig_Basilisk_setting} shows a time evolution of the interface positions of the film, with  the initial perturbations starting to grow due to the RP instability.
The axial velocity $u(x,r)$ is crucial to understand influence of the slip on the fibre thinning, shown in the lower half of figure\,\ref{fig_Basilisk_setting}, where opposite fluxes are found to be directed towards the left and right boundaries respectively, as the perturbation increases.
More velocity fields are presented in the left panel of figure\,\ref{fig_VF} for three different slip lengths at $t=100$.
According to the variations of the contours in the left panel of figure\,\ref{fig_VF}, the wall slip is found to accelerate the instability with larger $u(x,r)$. 
The velocity vectors in the lower half of figure\,\ref{fig_VF} show that slip decreases the velocity gradient $\partial u/ \partial r$ near the fibre.
The velocity profiles at different $x$-coordinates (see the dashed lines with arrows in the left panel of figure\,\ref{fig_VF}) are also extracted from the contours, shown in the right panel of figure\,\ref{fig_VF}, where the velocity on the fibre $u(x,\alpha)$ and its gradient $\partial_r u(x,\alpha)$ is calculated numerically.
According to (\ref{eq_stokes7}), $l_\mathrm{s} = u(x,\alpha) / \partial_r u(x,\alpha)$. So we can have numerically-predicted slip lengths at different positions, shown by the dashed lines in figure\,\ref{fig_VF}\,(d,f). These values agree with the input $l_{\mathrm{s}}$ in the boundary condition of the simulations, indicating that the numerical solutions can capture the flows features on the slip boundary conditions.

\begin{figure}
\centering
\includegraphics[width=1.0\textwidth]{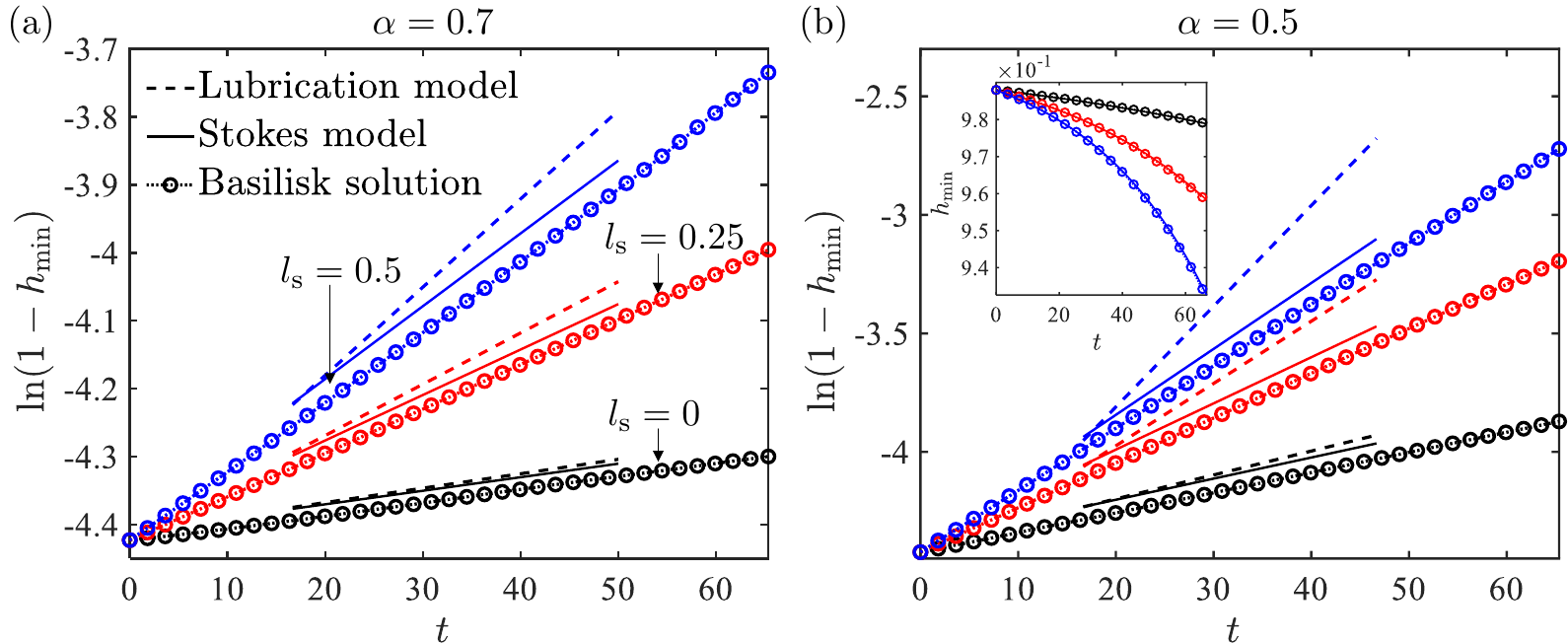}
	\caption{Linear time evolution of the minimum radii of films $h_\mathrm{min}(t)$ on two fibre radii: (a) $\alpha=0.7$; (b) $\alpha=0.5$. The numerical solutions (dotted lines with symbols) are compared to the predictions of both the lubrication model (dashed lines) and Stokes model (solid lines) for three slip lengths: $l_\mathrm{s} = 0$ (black), $0.25$ (red) and $0.5$ (blue). The inset illustrates $h_\mathrm{min}(t)$ in the uniform grids. Here $t$ is scaled by $\mu h_0/\gamma$.
	} 
\label{fig_hmint}	
\end{figure}

Figure\,\ref{fig_hmint} illustrates the growth rates of perturbations in the six cases simulated here.
Because $h(x,t) = 1+\hat{h}e^{ikx} e^{\omega t}$ is employed in the instability analysis (\S\,\ref{sec_instability}), we set the $y$-coordinate as $\mathrm{ln}(1-h_\mathrm{min})$ to present the linear growth of perturbations.
Here, the numerical solutions agree with predictions of the Stokes model, showing that the capability of the Stokes model to describe the RP instability of films on slippery fibres, while the predictions of the lubrication model are not accurate for the slip cases.
These results are consistent with dispersion relations in figure\,\ref{fig_dispersion2} and confirm the Stokes model numerically.

\begin{figure}
\centering
\includegraphics[width=1.0\textwidth]{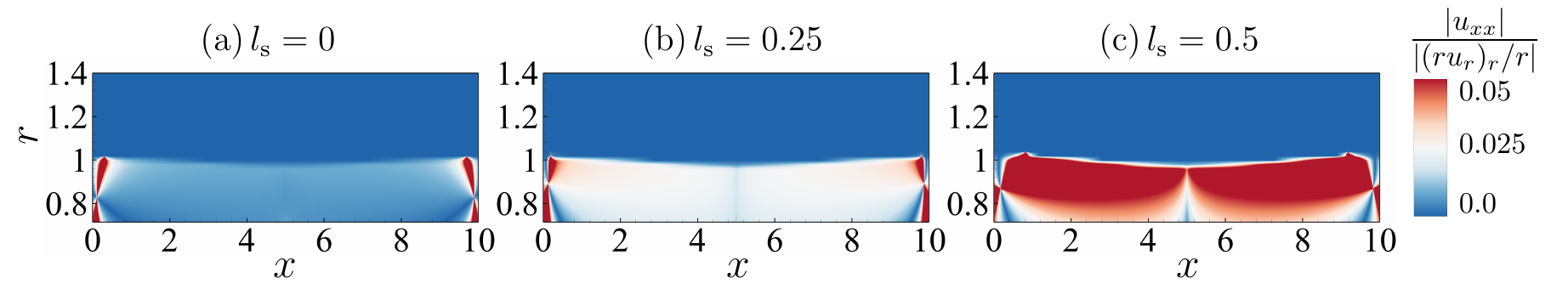}
	\caption{ Ratio of the velocity gradients at $t=100$ for different slip lengths: (a) $l_\mathrm{s} = 0$; (b) $l_\mathrm{s} = 0.25$; (c) $l_\mathrm{s} = 0.5$.  Here, the fibre radius $\alpha = 0.7$. $u_{xx}$ and $(r u_r)_r/r$ represents the second-order axial and radial derivative of the axial velocity in (\ref{eq_vel_deri}), respectively. }
\label{fig_vel_gradient}	
\end{figure}
Based on the approach proposed by \cite{liao2013drastic} and \cite{wei2019slipping}, we have a physical explanation for why large slip causes the failure of lubrication models, i.e. the radial derivative of the axial velocity is no longer much larger than its axial derivative, violating the lubrication approximation $\left| \frac{\partial^2 u}{\partial x^2} \right| \bigg/ \left|\frac{1}{r} \frac{\partial}{\partial r}\left(r \frac{\partial u}{\partial r} \right)\right| \ll 1 
 $ (see more details in Appendix\,\ref{app_LE_deri}). 
Figure\,\ref{fig_vel_gradient} shows the contour inside the film of the ratio $\left| \frac{\partial^2 u}{\partial x^2} \right| \bigg/ \left|\frac{1}{r} \frac{\partial}{\partial r}\left(r \frac{\partial u}{\partial r} \right)\right|$, extracted from the numerical solutions in this section. 
The ratio is found to increase from about $0.01$ of the no-slip case to $0.1$ of the slip case ($l_\mathrm{s}=0.5$) at the same location, which is consistent with the physical explanation above.

\subsection{Influence of slip on the dominant wavelength \label{subsec_dominant}}

For the influence of slip on the dominant wavelengths, we consider a configuration of a long film $L=160$ with $2^{13}$ grid points along the $x$-axis.
Eight different cases are simulated with four slip lengths, i.e. $l_\mathrm{s}=0$, $1$, $2$ and $3$ on fibres of two radii $\alpha=0.7$ and $\alpha=0.5$. 

\begin{figure}
\centering
\includegraphics[width=1.0\textwidth]{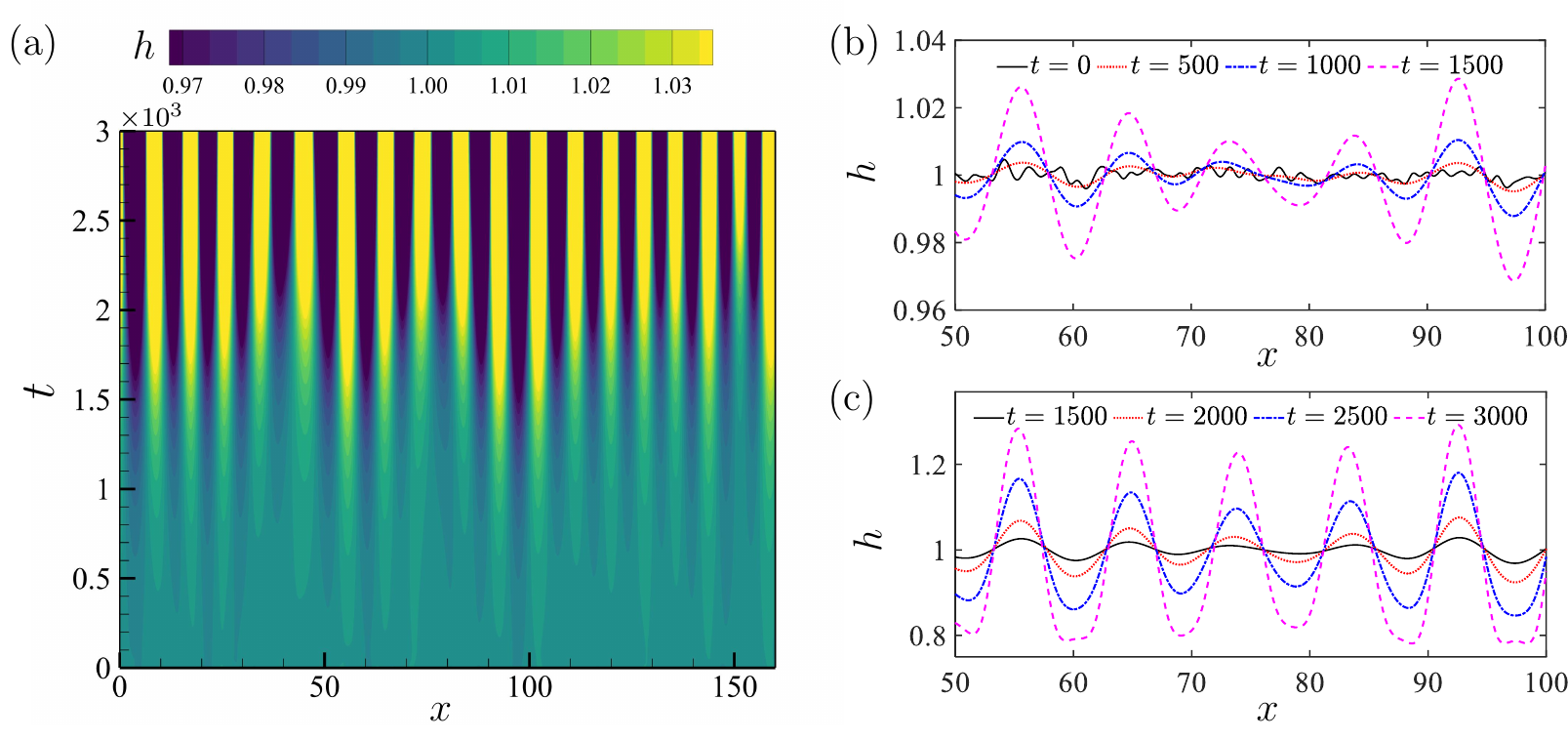}
	\caption{(a) Space-time plot of interface positions $h(x,t)$ on a no-slip fibre of the radius $\alpha=0.7$. Darker (brighter) color corresponds to the smaller (larger) values of $h(x,t)$. (b,c) Interface profiles extracted from (a) at early and late stages of the instability.}
\label{fig_LW}	
\end{figure}

The simulations start from random initial perturbations $h(x,0) = 1+ \varepsilon N(x)$, where $\varepsilon = 2 \times 10^{-3}$ and the random number $N(x)$ follows a normal distribution with zero mean and unit variance.
The random initial perturbations are designed to numerically mimic the arbitrary disturbances in the experiment \citep{haefner2015influence}.
Figure\,\ref{fig_LW} presents the spatial and temporal evolutions of interface profiles $h(x,t)$, where these small random perturbations, driven by the surface tension in the RP instability, grow gradually against time to generate significant capillary waves at early stage, plotted in  figure\,\ref{fig_LW}\,(b).
Figure\,\ref{fig_LW}\,(c) shows that wavelengths of these capillary waves are unchanged at the later stage of perturbation growth.

\begin{figure}
\centering
\includegraphics[width=0.65\textwidth]{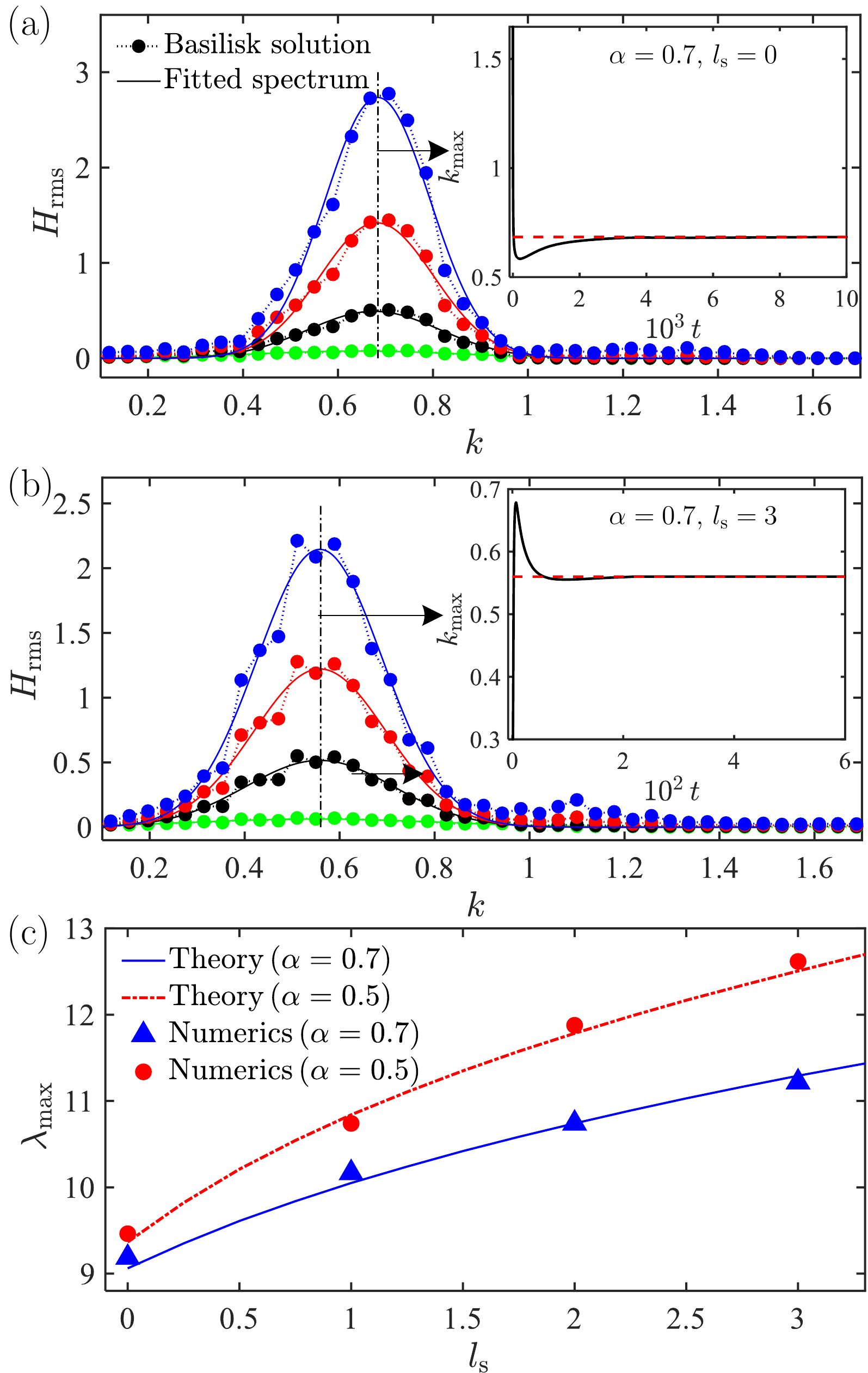}
	\caption{ (a,b) The root mean square (rms) of nondimensional perturbation amplitude versus nondimensional
wave number on fibres with different slip lengths at four time instants: (a) $2\times 10^3$ (green), $4.5\times 10^3$ (black), $6\times 10^3$ (red) and $7\times 10^3$ (blue); (b) $1.2\times 10^2$ (green), $3.3\times 10^2$ (black), $4.2\times 10^2$ (red) and $4.8\times 10^2$ (blue). The inset shows the time history of the dominant wavenumber.  
(c) Variations of the dominant wavelengths with slip lengths: a comparison between the theoretical predictions of (\ref{eq_dispersion_stokes}) and numerical solutions for two fibres of different radii. }
\label{fig_drop_statistic}	
\end{figure}

To further prove the previous findings and confirm theoretical predictions of the Stokes model quantitatively, multiple independent simulations (5 for each case) are performed to gather statistics of the dominant modes with a statistical methodology proposed by \cite{zhao2019revisiting}.
For each realisation, a discrete Fourier transform of the interface position $h(x)$ is applied to get the power spectral density (PSD) of the perturbations. 
The square root of the ensemble-averaged PSD ($H_\mathrm{rms}$) at each time is plotted in figure\,\ref{fig_drop_statistic}\,(a,b) with a modal distribution (spectrum) fitted by the Gaussian function, where the peak represents the dominant wavenumber $k_\mathrm{max}$ (see black dash-dotted lines).
Extracting $k_\mathrm{max}$ from the fitted spectrum at each time instant yields the insets in figure\,\ref{fig_drop_statistic}\,(a,b). 
Promisingly, $k_\mathrm{max}$ converges to a constant rapidly in both cases, further supporting the findings in figure\,\ref{fig_LW}\,(c).
Here, we compare these constants with the $k_\mathrm{max}$ predicted by the Stokes model.
Similar statistical analysis is applied for other cases to generate the symbols ($\lambda_\mathrm{max} = 2 \pi/k_\mathrm{max}$) in figure\,\ref{fig_drop_statistic}\,(c), giving good agreement with theoretical predictions.
Therefore, we can conclude that the dominant modes strongly depend on slip lengths with longer perturbation waves formed on a more slippery surface.

\begin{figure}
\centering
\includegraphics[width=0.65\textwidth]{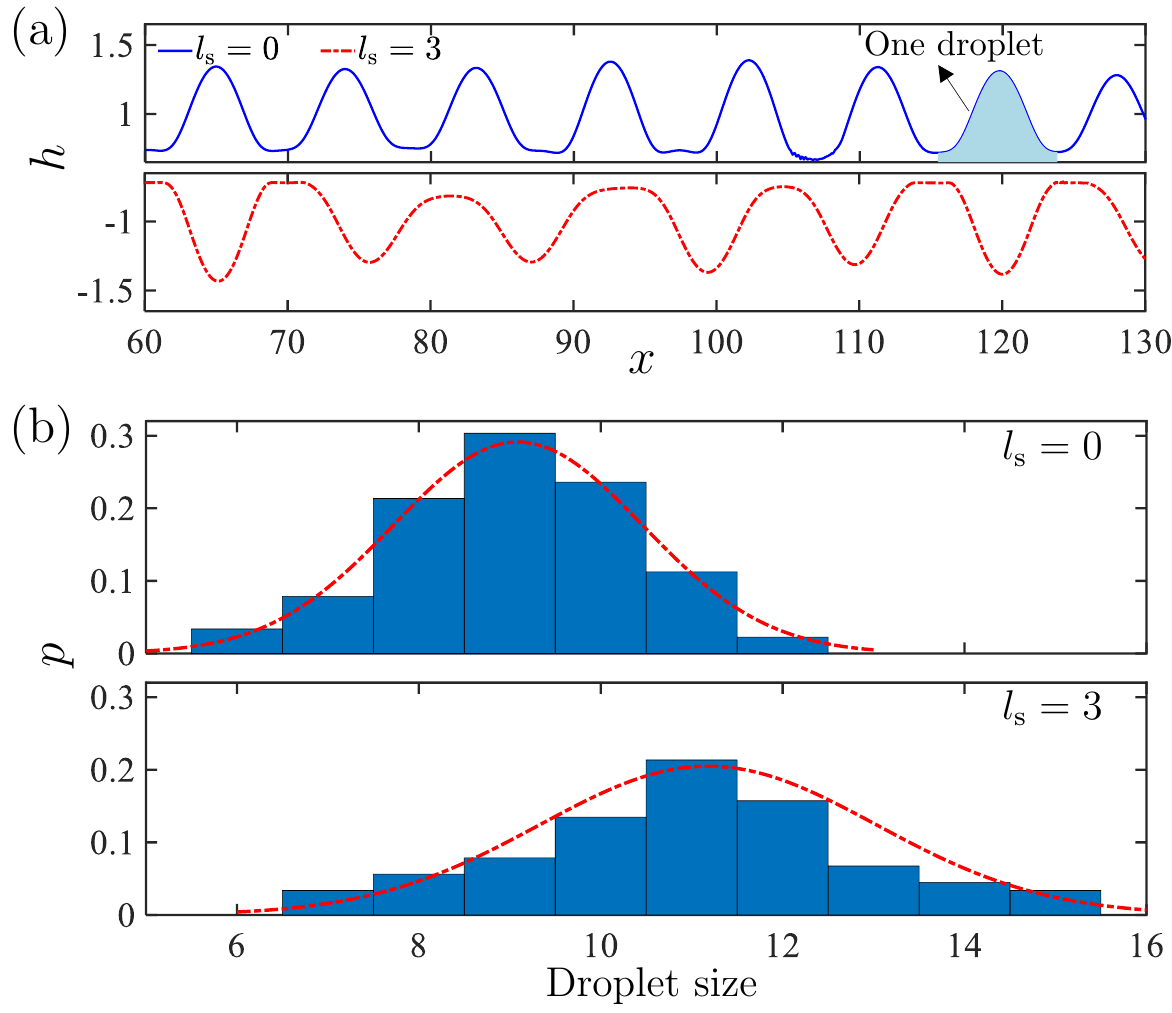}
	\caption{ (a) Interface profiles of liquid films on fibres of the radius $\alpha=0.7$. The results of two slip lengths are presented by solid blue lines for $l_\mathrm{s}=0$ and red dash-dotted lines for $l_\mathrm{s}=3$.
(b) Distributions of the droplet size on the no-slip (upper half) and slip (lower half).  	
	}
\label{fig_drop_size}	
\end{figure}

Figure\,{\ref{fig_drop_size}} shows the influence of slip on the droplet size.
In figure\,\ref{fig_drop_size}\,(a), larger drops are formed on slippery fibres with obviously longer wavelengths compared to ones achieved on the no-slip boundary, consistent with the theoretical results in figure\,\ref{fig_drop_statistic}\,(c) qualitatively.
Furthermore, we define the area between two bottom points of $h(x)$ as one droplet. 
Performing statistical analysis for all realisations gives the droplet-size distribution of a slip and no-slip case, illustrated in figure\,\ref{fig_drop_size}\,(b), whose mean and standard deviation is achieved by fitting the Gaussian function (red dash-dotted lines).
The slip case has a larger mean value of the distribution, further proving the finding in figure\,\ref{fig_drop_size}\,(a).
Besides the mean, slip is also found to lead to a wider distribution, with a larger standard deviation, possibly because the slip wall imposes less restrictions ( shear stresses on the fibre, i.e. $\tilde{\tau}_\mathrm{w} = \mu \partial_{\tilde{r}} \tilde{u}(\tilde{x},a) $) on the liquid compared to the no-slip one. 
The perturbation wavelengths have more `freedom' to expand or shrink, leading to a wider wavelength distribution.  
Note that figure\,\ref{fig_drop_size} might not be the final equilibrium state. So droplet coalescence could happen with complicated nonlinear dynamics. However, These nonlinear behaviours are beyond the scope of this work and should be the subject of future investigation.

\section{Comparisons with experimental results \label{sec_exp}}

Besides the numerical validations, the Stokes model is further compared with the experimental results in the previous work done by \cite{haefner2015influence}.
The experiments were done by measuring the evolutions of entangled polystyrene films with homogeneous initial thicknesses on no-slip and slip fibres of radius $a=9.6\,\mu$m.

\begin{figure}
\centering
\includegraphics[width=0.65\textwidth]{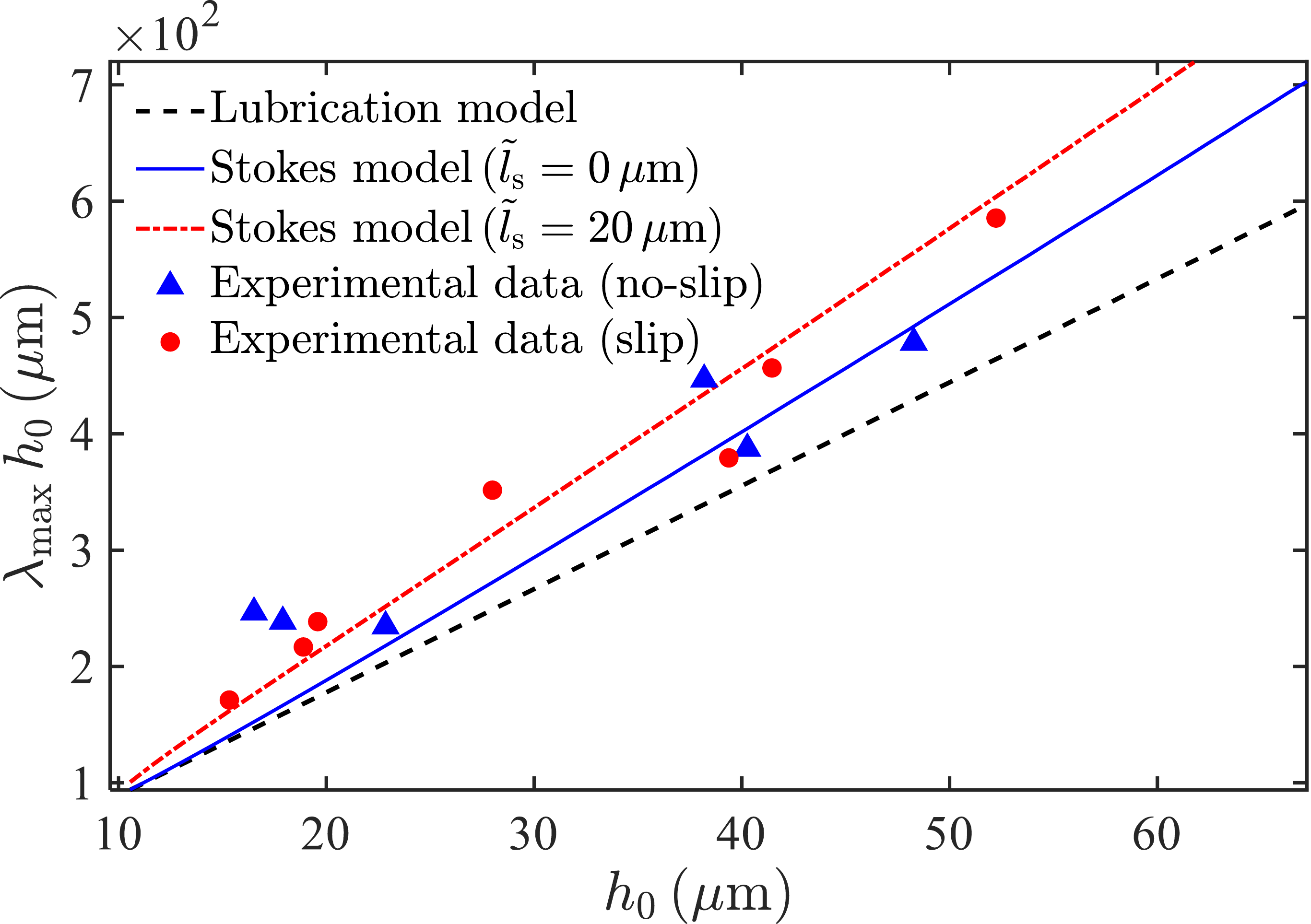}
	\caption{Influence of the geometry on the dominant wavelength of the instability: a comparison between the theoretical predictions and experimental data. The experimental data comes from the work of \cite{haefner2015influence}}.
\label{fig_experimental_data}	
\end{figure}

Figure\,\ref{fig_experimental_data} displays the dimensional dominant wavelengths $\tilde{\lambda}_\mathrm{max}$ ($\lambda_\mathrm{max} h_0$) of liquid films with different initial thickness $h_0$, where the blue triangles and red dots represent experimental data of the films on no-slip and slip fibres respectively.
Since $\lambda_\mathrm{max}$ predicted by the lubrication model (\ref{eq_dispersion_LE}) is the constant ($\lambda_\mathrm{max}=2\sqrt{2}\pi$), independent of both $l_\mathrm{s}$ and $\alpha$, $\lambda_\mathrm{max} h_0$ grows linearly with increasing $h_0$ (see the black dashed lines in figure\,\ref{fig_experimental_data}).
However, the lubrication model seemingly underestimates experimentally-obtained $\lambda_\mathrm{max}$.
The possible reason is that most $\alpha$ values investigated in the experiment are smaller than $0.5$, outside the range of validity of the lubrication approximation.
A series of dominant wavelengths with different $\alpha$ values are extracted from the Stokes model (\ref{eq_dispersion_stokes}) to generate a blue solid line for no-slip cases and a red dash-dotted line for slip cases with a fitting $\tilde{l}_\mathrm{s}$ in figure\,\ref{fig_experimental_data}.
Note that the dimensionless slip length varies along the red dash-dotted curve with different $h_0$ values.
Though both the blue and red curves still look linear due to small changes of $\lambda_\mathrm{max}$, obvious differences between the Stokes predictions and the lubrication ones are found for the thick films ($h_0>40\,\mu$m).
Promisingly, despite dispersion of experimental data, the Stokes model can provide better predictions for dominant wavelengths of both no-slip cases and slip cases compared to the lubrication model.
%denoting that the new model is a powerful tool to describe the slip-enhanced RP instability of a film on a fibre.

\begin{figure}
\centering
\includegraphics[width=0.65\textwidth]{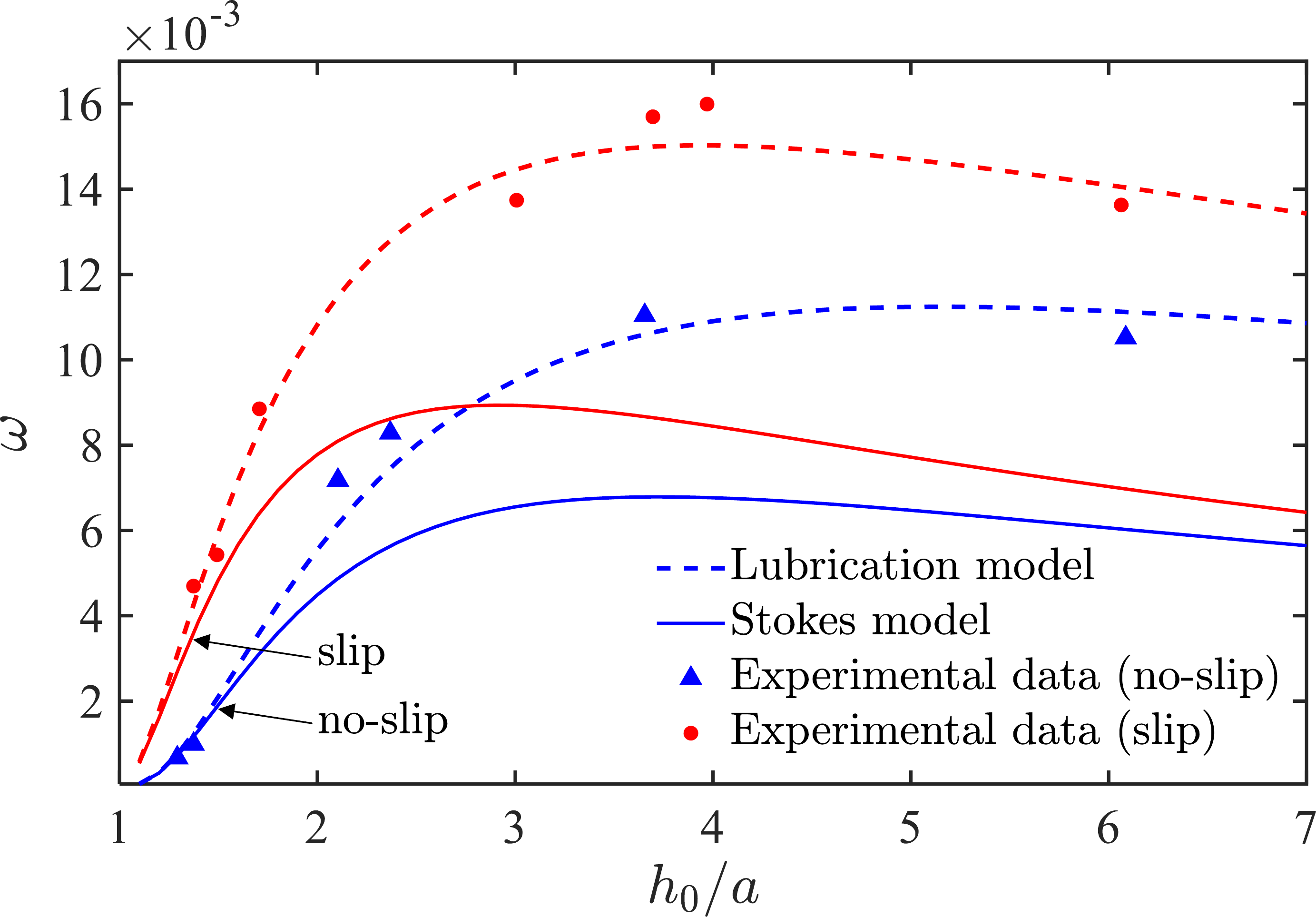}
	\caption{Influence of the geometry on the growth rate of the dominant mode on no-slip and slip fibres. The solid lines and dashed lines are the predictions of the Stokes model and the lubrication model, respectively. Blue lines represent no-slip results and red lines are the slip ones. The experimental data comes from the work of \cite{haefner2015influence}. }
\label{fig_experimental_data2}	
\end{figure}

In addition to the dominant wavelengths, we then make a further comparison between the experimental growth rates and theoretical ones, illustrated in figure\,\ref{fig_experimental_data2}, where the growth rate is non-dimensionalised by the ratio of the capillary velocity to the fibre radius, i.e., $\omega  =\frac{a}{\gamma/\mu }  \tilde{\omega}$.  
However, using the parameters from \cite{haefner2015influence} ($\gamma/\mu = 294\,\mu\mathrm{m\,min}^{-1}$ and $\tilde{l}_\mathrm{s} = 0.3\, a$), the Stokes model is found to provide worse agreements with the experimental data than the lubrication model.
Noticeably, predictions of the lubrication model match the experimental data even in the cases of thick films ($h_0/a > 3$), which is unreasonable due to the thin-film assumption in the lubrication theory. 
The reason for the disagreement is currently unclear and should be the subject of future investigation. 

\section{Conclusions}  
In this work, to study the influence of liquid-sold slip on the RP instability of liquid films on a fibre, a theoretical model based on the linear stability analysis of axisymmetric Stokes equations is developed, which goes beyond the classical slip lubrication model. The lubrication model is found to overestimate the slip enhanced growth rates of perturbations and fail to predict the dominant wavelength, while the new model can provide more accurate predictions for the instability. 
Direct numerical simulations of the NS equations are then performed to validate the two theoretical findings via two fluid configurations: (i) a short film with a fixed wavelength for the slip-enhanced growth rate of the perturbation, and (ii) a long film with random initial perturbations for the slip-dependent dominant modes of perturbations. 
The dominant wavelengths are also confirmed by experimental data from the previous work \citep{haefner2015influence}.
Overall, the applicability of the slip lubrication model should be restricted to thin-film and small-slip flows. For more general problems, it is better to use the Stokes theory proposed here.

Promisingly, the wall slip has been realised experimentally for films on a fibre \citep{haefner2015influence,ji2019dynamics} and the slip length can also be measured directly \citep{huang2006direct,maali2012measurement,maali2016slip}.
It is our hope that the wall slip can be controlled
experimentally to verify our new predictions, most notably the relation between drop sizes ($\lambda_\mathrm{max}$) and slip lengths. 
Potential extensions of the framework are numerous.
For example, to include the influence of other physics such as electric fields \citep{ding2014dynamics} and intermolecular forces \citep{ji2019dynamics,tomo2022observation}, both of which are know to affect $k_\mathrm{crit}$ and $k_\mathrm{max}$ in the RP instability, and to explore related flow configurations, like a film flowing down a fibre \citep{liu2021coating}, where an interesting open problem is to understand the influence of slip on dynamics of travelling waves in different flow regimes.

%\section*{Acknowledgements}~~~~~~~~~~~~~~~~~~~~~~~~~~~~~~~
\backsection[Acknowledgements]{
Useful discussions with Dr. Mykyta Chubynsky, Dr. Mu Kai, Mr. Qiao Ran and Dr. Qin Jian are gratefully acknowledged.}

%funding
\backsection[Funding]{This work was supported by the National Natural Science Foundation of China (grant number $11621202$) and the Youth Innovation Promotion Association CAS (grant number $2018491$).} 

%\section*{Declaration of interests}~~~~~~~~~~~~~~~~~~~~~~~
\backsection[Declaration of interests]{
The authors report no conflict of interest.}

\backsection[Author ORCIDs]{\\
Chengxi Zhao https://orcid.org/0000-0002-3041-0882\\
Yixin Zhang https://orcid.org/0000-0003-4632-3780\\
Ting Si https://orcid.org/0000-0001-9071-8646
}

\appendix
\section{Derivation for the lubrication equation }\label{app_LE_deri}
In this appendix, we present a detailed reproduction for the derivation of the lubrication equation used by \cite{haefner2015influence}. 
To get this lubrication equation from the axisymmetric Navier-Stokes equations, we need to establish the leading order terms by their asymptotic expansion in $\varepsilon$, for which we use the rescaling shown below: 
$$
\tilde{x}= \tilde{\lambda} x,\,\, \tilde{r}=\varepsilon \tilde{\lambda} r,\,\, 
\tilde{u} = \varepsilon^3 \frac{\gamma}{\mu} u,\,\,\tilde{v}=\varepsilon^4\frac{\gamma}{\mu} v,\,\,
\tilde{t} = \frac{ \tilde{\lambda} \mu }{\varepsilon^3 \gamma} t,\,\,
\tilde{p} = \frac{\varepsilon \gamma }{ \tilde{\lambda}} p\,.
$$
Here, $ \tilde{\lambda} = h_0 / \varepsilon  $. 
Substituting all these scalings into the dimensional Stokes equations yields,
 \begin{equation}
\frac{\partial u}{\partial x} + \frac{1}{r} \frac{\partial(vr) }{\partial r}  = 0\,, \\
 \end{equation}
  \begin{equation}
  \label{eq_app_momentum}
 \frac{\partial p }{\partial x} = \varepsilon^2 \frac{\partial^2 u}{\partial x^2} 
  + \frac{1}{r} \frac{\partial}{\partial r}\left(r \frac{\partial u}{\partial r} \right)  \,,
  \end{equation}
    \begin{equation}
    \frac{\partial p}{\partial r} =
    \varepsilon^4 \frac{\partial^2 v}{\partial x^2} 
    +\varepsilon^2 \frac{\partial}{\partial r} \left[\frac{1}{r} \frac{\partial (vr)}{\partial r}\right]  \,. 
  \end{equation}
The equations at the liquid-gas interface ($r=h$) are scaled as
  \begin{equation}
     \frac{\partial h}{\partial t} + u \frac{\partial h}{\partial x} -v = 0 \,,
\end{equation}     
  \begin{equation}
  \label{eq_scaling_5}
  \begin{split}
  p - \frac{2 \varepsilon^2}{1+ \varepsilon^2 \left(\partial_x h \right)^2} 
  \left[ \frac{\partial v}{\partial r} - \frac{\partial h}{\partial x} \left(  \frac{\partial u}{\partial r} 
  + \varepsilon^2 \frac{\partial v}{\partial x}\right) + \varepsilon^2 \left(\frac{\partial h}{\partial x} \right)^2 \frac{\partial u}{\partial x} \right]  \\
 =\frac{1}{\varepsilon^2 h \sqrt{1+ \varepsilon^2 \left(\partial_x h \right)^2}}  - \frac{\partial_x^2 h}{\left(1+\varepsilon^2 \left(\partial_x h \right)^2\right) ^\frac{3}{2}}   \,,
  \end{split}
  \end{equation} 
  \begin{equation}
  \label{eq_scaling_6}
  2 \varepsilon^2 \frac{\partial h}{\partial x} \left( \frac{\partial v}{\partial r} - \frac{\partial u}{\partial x} \right) 
  + \left[ 1- \varepsilon^2 \left(\frac{\partial h}{\partial x} \right)^2 \right]  \left( \frac{\partial u}{\partial r} + \varepsilon^2 \frac{\partial v}{\partial x} \right) = 0 \,.
  \end{equation}
For the boundary conditions at the liquid-solid interface ($r=\alpha$) the scaled form are
\begin{align}
\label{eq_scaling_7}
 & u = \frac{\tilde{l}_\mathrm{s}}{ \varepsilon \tilde{\lambda}} \frac{\partial u}{\partial r},\,\, v=0 \,.
\end{align}

After eliminating all the high order terms of $\varepsilon$, we obtain
 \begin{equation}
  \label{eq_rescale_mass}
 \partial_x u+ \partial_r(vr)/r = 0 \,, 
 \end{equation}
  \begin{equation}
    \label{eq_rescale_x_momen}
  0= - \partial_x p+ \partial_r \left(r \partial_r u \right)/r \,, 
  \end{equation}
    \begin{equation}
    \label{eq_rescale_y_momen}
  0 =  \partial_r p \,,
  \end{equation}
  \begin{equation}
      \label{eq_rescale_kine}
     \partial_{t} h  + u\, \partial_{x} h  -v = 0 \,,
\end{equation}     
\begin{equation}
\label{eq_rescale_nor}
 p  =  1/h  - \partial_{x}^2 h,  \quad (r=h) \,,
 \end{equation}
\begin{equation}
\label{eq_rescale_tange}
 \partial_r u = 0, \quad (r=h) \,,
\end{equation}
\begin{equation}
\label{eq_rescale_slip}
 u = l_\mathrm{s} \partial_r u, \quad (r=\alpha) \,,
\end{equation}
\begin{equation}
\label{eq_rescale_no_pen}
 v = 0,  \quad (r=\alpha) \,.
\end{equation}

Notably, the term $\partial_x^2 h$ is not in the leading order compared to $1/h$ in (\ref{eq_rescale_nor}), but conventionally in this field, it is still included in the pressure in an attempt to extend the validity of the model \citep{eggers1994drop, craster2006viscous}. 
Additionally, (\ref{eq_scaling_7}) shows that the leading-order slip boundary depends on the ratio $\tilde{l}_\mathrm{s}/h_0$. 
In this work, we consider the slip is at the order of $h_0$ so that the scaled slip boundary condition is
unchanged in (\ref{eq_rescale_slip}).

Integrating equation\,(\ref{eq_rescale_x_momen}) from $r=h$ to $r=r$ with the boundary condition equation\,(\ref{eq_rescale_tange}) give us
\begin{equation}
\frac{1}{2} \frac{\partial p}{\partial x} \left( r^2 - h^2\right) = r\frac{\partial u}{\partial r}.
\label{eq_LE_p1}
\end{equation}
After integrating (\ref{eq_LE_p1}) from $r=\alpha$ to $r=r$ with the slip boundary condition (\ref{eq_rescale_slip}), we have
\begin{equation}
\frac{1}{2} \frac{\partial p}{\partial x} \left[ \frac{1}{2} (r^2 - \alpha^2) - h^2\, \mathrm{ln}(r/\alpha)\right] = u- l_\mathrm{s}\frac{\partial u}{\partial r}|_{r=\alpha}.
\label{eq_LE_p2}
\end{equation}
Combine (\ref{eq_LE_p1}) and (\ref{eq_LE_p2}) yields
\begin{equation}
\label{eq_LE_velocity}
u = \frac{1}{2} \frac{\partial p}{\partial x} \left[ \frac{1}{2} (r^2 - \alpha^2) - h^2\, \mathrm{ln}\left(\frac{r}{\alpha} \right) + l_\mathrm{s}\left(\alpha-\frac{h^2}{\alpha} \right) \right] \,.
\end{equation}
Using equation\,(\ref{eq_rescale_mass}), (\ref{eq_rescale_kine}) and (\ref{eq_rescale_no_pen}) with the Leibniz integral rule, one can obtain
\begin{equation}
h\frac{\partial h}{\partial t} = -\frac{\partial}{\partial x} \int_\alpha^h ur dr
\end{equation}
Substituting equation\,(\ref{eq_LE_velocity}) into the integral results in
\begin{equation}
\label{eq_LE_frame_app}
\frac{\partial h}{\partial t} = \frac{1}{h} \frac{\partial}{\partial x} \left[ M(h) \frac{\partial p}{\partial x} \right]\,,
\end{equation}
where
$$M(h) = \left[-3h^4-\alpha^4 + 4 \alpha^2 h^2 + 4 h^4 \mathrm{ln}\left(h/\alpha\right) + 4 l_\mathrm{s} (h^2-\alpha^2)^2 / \alpha \right] / 16,$$
and
$$ p  = 1/h  - \partial_x^2 h \,. $$

Note that this lubrication equation (\ref{eq_LE_frame_app}) is not available for a large-slip case because $\tilde{l}_\mathrm{s} \sim \varepsilon \tilde{\lambda}$, derived from (\ref{eq_scaling_7}).
Moreover, $\tilde{l}_\mathrm{s}$ was found not to be arbitrarily large but has to be under a certain constraint to make the lubrication approximation hold \citep{liao2013drastic,wei2019slipping}.
To derive such constraint for (\ref{eq_LE_frame_app}), we start from (\ref{eq_app_momentum}), where the radial derivative of the axial velocity is found to be much larger than its axial derivative, 
\begin{equation}
\label{eq_vel_deri}
\left|\frac{1}{\tilde{r}} \frac{\partial}{\partial \tilde{r}}\left(\tilde{r} \frac{\partial \tilde{u}}{\partial \tilde{r}} \right)\right| \gg 
\left| \frac{\partial^2 \tilde{u}}{\partial \tilde{x}^2} \right|. 
\end{equation}  
 Expanding the radial derivative yields 
 \begin{equation}
 \label{eq_vel_deri_scale}
 \left|\frac{1}{\tilde{r}} \frac{\partial \tilde{u}}{\partial \tilde{r}}\right| \gg 
\left| \frac{\partial^2 \tilde{u}}{\partial \tilde{x}^2} \right|
\textrm{ or }  \left|\frac{\partial^2 \tilde{u}}{\partial \tilde{r}^2}\right| \gg 
\left| \frac{\partial^2 \tilde{u}}{\partial \tilde{x}^2} \right|.
 \end{equation}
With a similar approach employed by \cite{liao2013drastic} and \cite{wei2019slipping}, the radial derivative is scaled as
\begin{align}
\label{eq_vel_scale1}
& \frac{1}{\tilde{r}} \frac{\partial \tilde{u}}{\partial \tilde{r}} \sim  \frac{\triangle U}{h_0 (h_0-a)} \sim  \frac{U}{h_0 (h_0-a+\tilde{l}_\mathrm{s})}\,,  \\
& \frac{\partial^2 \tilde{u}}{\partial \tilde{r}^2}  \sim  \frac{\triangle U}{(h_0-a)^2} \sim \frac{U}{(h_0-a)(h_0-a+\tilde{l}_\mathrm{s})}\,, 
\end{align}
where $U$ is the dimensional characteristic velocity and $\triangle U \sim \frac{h_0-a}{h_0-a+\tilde{l}_\mathrm{s}} U$.
The axial one is
\begin{equation}
\label{eq_vel_scale3}
\partial^2_{\tilde{x}} \tilde{u}  \sim U/ \tilde{\lambda}^2 \,.
\end{equation}
Substituting (\ref{eq_vel_scale1})-(\ref{eq_vel_scale3}) into (\ref{eq_vel_deri_scale}) gives the final constraint, 
\begin{equation}
\tilde{l}_\mathrm{s} \ll \tilde{\lambda}^2/h_0 - h_0+a\,,
\end{equation}
 whose dimensionless format is
 \begin{equation}
 \label{eq_ls_constraint}
l_\mathrm{s} \ll \lambda^2 -1+\alpha\,,
\end{equation}

\bibliographystyle{jfm}
\bibliography{jfm}

\begin{thebibliography}{56}
\expandafter\ifx\csname natexlab\endcsname\relax\def\natexlab#1{#1}\fi
\def\au#1{#1} \def\ed#1{#1} \def\yr#1{#1}\def\at#1{#1}\def\jt#1{\textit{#1}}
  \def\bt#1{#1}\def\bvol#1{\textbf{#1}} \def\vol#1{#1} \def\pg#1{#1}
  \def\publ#1{#1}\def\arxiv#1{#1}\def\org#1{#1}\def\st#1{\textit{#1}}

\bibitem[Berny {\em et~al.\/}(2020)Berny, Deike, S{\'e}on \&
  Popinet]{berny2020role}
{\sc \au{Berny, A.}, \au{Deike, L.}, \au{S{\'e}on, T.} \& \au{Popinet, S.}}
  \yr{2020}  \at{Role of all jet drops in mass transfer from bursting bubbles}.
   \jt{Phys. Rev. Fluids}  \bvol{5}~(3),  \pg{033605}.

\bibitem[Brochard-Wyart {\em et~al.\/}(1994)Brochard-Wyart, De~Gennes, Hervert
  \& Redon]{brochard1994wetting}
{\sc \au{Brochard-Wyart, F.}, \au{De~Gennes, P.~G.}, \au{Hervert, H.} \&
  \au{Redon, C.}} \yr{1994}  \at{Wetting and slippage of polymer melts on
  semi--ideal surfaces}.  \jt{Langmuir}  \bvol{10}~(5),  \pg{1566--1572}.

\bibitem[Chao {\em et~al.\/}(2018)Chao, Ding \& Liu]{chao2018dynamics}
{\sc \au{Chao, Y.}, \au{Ding, Z.} \& \au{Liu, R.}} \yr{2018}  \at{Dynamics of
  thin liquid films flowing down the uniformly heated/cooled cylinder with wall
  slippage}.  \jt{Chem. Eng. Sci.}  \bvol{175},  \pg{354--364}.

\bibitem[Chen {\em et~al.\/}(2018)Chen, Ran, Gan, Zhou, Zhang, Zhang, Zhang \&
  Jiang]{chen2018ultrafast}
{\sc \au{Chen, H.}, \au{Ran, T.}, \au{Gan, Y.}, \au{Zhou, J.}, \au{Zhang, Y.},
  \au{Zhang, L.}, \au{Zhang, D.} \& \au{Jiang, L.}} \yr{2018}  \at{Ultrafast
  water harvesting and transport in hierarchical microchannels}.  \jt{Nat.
  Mater.}  \bvol{17}~(10),  \pg{935--942}.

\bibitem[Craster \& Matar(2006)]{craster2006viscous}
{\sc \au{Craster, R.~V.} \& \au{Matar, O.~K.}} \yr{2006}  \at{On viscous beads
  flowing down a vertical fibre}.  \jt{J. Fluid Mech.}  \bvol{553},
  \pg{85--105}.

\bibitem[Craster \& Matar(2009)]{craster2009dynamics}
{\sc \au{Craster, R.~V.} \& \au{Matar, O.~K.}} \yr{2009}  \at{Dynamics and
  stability of thin liquid films}.  \jt{Rev. Mod. Phys}  \bvol{81}~(3),
  \pg{1131}.

\bibitem[Deblais {\em et~al.\/}(2018)Deblais, Herrada, Hauner, Velikov,
  Van~Roon, Kellay, Eggers \& Bonn]{deblais2018viscous}
{\sc \au{Deblais, A.}, \au{Herrada, M.~A.}, \au{Hauner, I.}, \au{Velikov,
  K.~P.}, \au{Van~Roon, T.}, \au{Kellay, H.}, \au{Eggers, J.} \& \au{Bonn, D.}}
  \yr{2018}  \at{Viscous effects on inertial drop formation}.  \jt{Phys. Rev.
  Lett.}  \bvol{121}~(25),  \pg{254501}.

\bibitem[Deng {\em et~al.\/}(2011)Deng, Nave, Liang, Johnson \&
  Fink]{deng2011exploration}
{\sc \au{Deng, D.}, \au{Nave, J.}, \au{Liang, X.}, \au{Johnson, S.~G.} \&
  \au{Fink, Y.}} \yr{2011}  \at{Exploration of in-fiber nanostructures from
  capillary instability}.  \jt{Opt. Express}  \bvol{19}~(17),
  \pg{16273--16290}.

\bibitem[Ding \& Liu(2011)]{ding2011stability}
{\sc \au{Ding, Z.} \& \au{Liu, Q.}} \yr{2011}  \at{Stability of liquid films on
  a porous vertical cylinder}.  \jt{Phys. Rev. E}  \bvol{84}~(4),  \pg{046307}.

\bibitem[Ding {\em et~al.\/}(2013)Ding, Wong, Liu \& Liu]{ding2013viscous}
{\sc \au{Ding, Z.}, \au{Wong, T.~N.}, \au{Liu, R.} \& \au{Liu, Q.}} \yr{2013}
  \at{Viscous liquid films on a porous vertical cylinder: Dynamics and
  stability}.  \jt{Phys. Fluids}  \bvol{25}~(6),  \pg{064101}.

\bibitem[Ding {\em et~al.\/}(2014)Ding, Xie, Wong \& Liu]{ding2014dynamics}
{\sc \au{Ding, Z.}, \au{Xie, J.}, \au{Wong, T.~N.} \& \au{Liu, R.}} \yr{2014}
  \at{Dynamics of liquid films on vertical fibres in a radial electric field}.
  \jt{J. Fluid Mech.}  \bvol{752},  \pg{66--89}.

\bibitem[Duprat {\em et~al.\/}(2007)Duprat, Ruyer-Quil, Kalliadasis \&
  Giorgiutti-Dauphin{\'e}]{duprat2007absolute}
{\sc \au{Duprat, C.}, \au{Ruyer-Quil, C.}, \au{Kalliadasis, S.} \&
  \au{Giorgiutti-Dauphin{\'e}, F.}} \yr{2007}  \at{Absolute and convective
  instabilities of a viscous film flowing down a vertical fiber}.  \jt{Phys.
  Rev. Lett.}  \bvol{98}~(24),  \pg{244502}.

\bibitem[Eggers \& Dupont(1994)]{eggers1994drop}
{\sc \au{Eggers, J.} \& \au{Dupont, T.~F.}} \yr{1994}  \at{Drop formation in a
  one--dimensional approximation of the navier--stokes equation}.  \jt{J. Fluid
  Mech.}  \bvol{262},  \pg{205--221}.

\bibitem[Frenkel(1992)]{frenkel1992nonlinear}
{\sc \au{Frenkel, A.~L.}} \yr{1992}  \at{Nonlinear theory of strongly
  undulating thin films flowing down vertical cylinders}.  \jt{Europhys. Lett.}
   \bvol{18}~(7),  \pg{583}.

\bibitem[Goren(1962)]{goren1962instability}
{\sc \au{Goren, S.~L.}} \yr{1962}  \at{The instability of an annular thread of
  fluid}.  \jt{J. Fluid Mech.}  \bvol{12}~(2),  \pg{309--319}.

\bibitem[Goren(1964)]{goren1964shape}
{\sc \au{Goren, S.~L.}} \yr{1964}  \at{The shape of a thread of liquid
  undergoing break--up}.  \jt{J. Colloid Sci.}  \bvol{19}~(1),  \pg{81--86}.

\bibitem[Haefner(2015)]{haefner2015rayleigh}
{\sc \au{Haefner, S.}} \yr{2015}  \at{{R}ayleigh--{P}lateau--type instabilities
  in thin liquid films}. PhD thesis, Saarland University.

\bibitem[Haefner {\em et~al.\/}(2015)Haefner, Benzaquen, B{\"a}umchen, Salez,
  Peters, McGraw, Jacobs, Rapha{\"e}l \& Dalnoki-Veress]{haefner2015influence}
{\sc \au{Haefner, S.}, \au{Benzaquen, M.}, \au{B{\"a}umchen, O.}, \au{Salez,
  T.}, \au{Peters, R.}, \au{McGraw, J.~D}, \au{Jacobs, K.}, \au{Rapha{\"e}l,
  E.} \& \au{Dalnoki-Veress, K.}} \yr{2015}  \at{Influence of slip on the
  {P}lateau--{R}ayleigh instability on a fibre}.  \jt{Nat. Commun.}
  \bvol{6}~(1),  \pg{1--6}.

\bibitem[Halpern {\em et~al.\/}(2015)Halpern, Li \& Wei]{halpern2015slip}
{\sc \au{Halpern, D.}, \au{Li, Y.} \& \au{Wei, H.}} \yr{2015}
  \at{Slip--induced suppression of marangoni film thickening in
  surfactant--retarded {L}andau--{L}evich--{B}retherton flows}.  \jt{J. Fluid
  Mech.}  \bvol{781},  \pg{578--594}.

\bibitem[Halpern \& Wei(2017)]{halpern2017slip}
{\sc \au{Halpern, D.} \& \au{Wei, H.}} \yr{2017}  \at{Slip--enhanced drop
  formation in a liquid falling down a vertical fibre}.  \jt{J. Fluid Mech.}
  \bvol{820},  \pg{42--60}.

\bibitem[Hammond(1983)]{hammond1983nonlinear}
{\sc \au{Hammond, P.~S.}} \yr{1983}  \at{Nonlinear adjustment of a thin annular
  film of viscous fluid surrounding a thread of another within a circular
  cylindrical pipe}.  \jt{J. Fluid Mech.}  \bvol{137},  \pg{363--384}.

\bibitem[Huang {\em et~al.\/}(2006)Huang, Guasto \& Breuer]{huang2006direct}
{\sc \au{Huang, P.}, \au{Guasto, J.~S.} \& \au{Breuer, K.~S.}} \yr{2006}
  \at{Direct measurement of slip velocities using three--dimensional total
  internal reflection velocimetry}.  \jt{J. Fluid Mech.}  \bvol{566},
  \pg{447--464}.

\bibitem[Ji {\em et~al.\/}(2019)Ji, Falcon, Sadeghpour, Zeng, Ju \&
  Bertozzi]{ji2019dynamics}
{\sc \au{Ji, H.}, \au{Falcon, C.}, \au{Sadeghpour, A.}, \au{Zeng, Z.}, \au{Ju,
  Y.~S.} \& \au{Bertozzi, A.~L.}} \yr{2019}  \at{Dynamics of thin liquid films
  on vertical cylindrical fibres}.  \jt{J. Fluid Mech.}  \bvol{865},
  \pg{303--327}.

\bibitem[Kalliadasis \& Chang(1994)]{kalliadasis1994drop}
{\sc \au{Kalliadasis, S.} \& \au{Chang, H.}} \yr{1994}  \at{Drop formation
  during coating of vertical fibres}.  \jt{J. Fluid Mech.}  \bvol{261},
  \pg{135--168}.

\bibitem[Kavokine {\em et~al.\/}(2022)Kavokine, Bocquet \&
  Bocquet]{kavokine2022fluctuation}
{\sc \au{Kavokine, N.}, \au{Bocquet, M.} \& \au{Bocquet, L.}} \yr{2022}
  \at{Fluctuation--induced quantum friction in nanoscale water flows}.
  \jt{Nature}  \bvol{602}~(7895),  \pg{84--90}.

\bibitem[Kavokine {\em et~al.\/}(2021)Kavokine, Netz \&
  Bocquet]{kavokine2021fluids}
{\sc \au{Kavokine, N.}, \au{Netz, R.~R.} \& \au{Bocquet, L.}} \yr{2021}
  \at{Fluids at the nanoscale: From continuum to subcontinuum transport}.
  \jt{Annu. Rev. Fluid Mech.}  \bvol{53},  \pg{377--410}.

\bibitem[Kliakhandler {\em et~al.\/}(2001)Kliakhandler, Davis \&
  Bankoff]{kliakhandler2001viscous}
{\sc \au{Kliakhandler, I.~L.}, \au{Davis, S.~H.} \& \au{Bankoff, S.~G.}}
  \yr{2001}  \at{Viscous beads on vertical fibre}.  \jt{J. Fluid Mech.}
  \bvol{429},  \pg{381--390}.

\bibitem[Lauga {\em et~al.\/}(2007)Lauga, Brenner \& Stone]{Lauga2007}
{\sc \au{Lauga, E.}, \au{Brenner, M.} \& \au{Stone, H.}} \yr{2007} {\em
  Microfluidics: The No--Slip Boundary Condition\/},  \pg{pp. 1219--1240}.
  \publ{Berlin, Heidelberg: Springer}.

\bibitem[Lee {\em et~al.\/}(2022)Lee, Chan, Carlson \&
  Dalnoki-Veress]{lee2022multiple}
{\sc \au{Lee, C.~L.}, \au{Chan, T.~S.}, \au{Carlson, A.} \& \au{Dalnoki-Veress,
  K.}} \yr{2022}  \at{Multiple droplets on a conical fiber: formation, motion,
  and droplet mergers}.  \jt{Soft Matter}  \bvol{18},  \pg{134495}.

\bibitem[Liang {\em et~al.\/}(2011)Liang, Deng, Nave \&
  Johnson]{liang2011linear}
{\sc \au{Liang, X.}, \au{Deng, D.}, \au{Nave, J.} \& \au{Johnson, S.~G.}}
  \yr{2011}  \at{Linear stability analysis of capillary instabilities for
  concentric cylindrical shells}.  \jt{J. Fluid Mech.}  \bvol{683},
  \pg{235--262}.

\bibitem[Liao {\em et~al.\/}(2014)Liao, Li, Chang, Huang \&
  Wei]{liao2014speeding}
{\sc \au{Liao, Y.}, \au{Li, Y.}, \au{Chang, Y.}, \au{Huang, C.} \& \au{Wei,
  H.}} \yr{2014}  \at{Speeding up thermocapillary migration of a confined
  bubble by wall slip}.  \jt{J. Fluid Mech.}  \bvol{746},  \pg{31--52}.

\bibitem[Liao {\em et~al.\/}(2013)Liao, Li \& Wei]{liao2013drastic}
{\sc \au{Liao, Y.}, \au{Li, Y.} \& \au{Wei, H.}} \yr{2013}  \at{Drastic changes
  in interfacial hydrodynamics due to wall slippage: slip-intensified film
  thinning, drop spreading, and capillary instability}.  \jt{Phys. Rev. Lett.}
  \bvol{111}~(18),  \pg{1364--1370}.

\bibitem[Liu \& Ding(2021)]{liu2021coating}
{\sc \au{Liu, R.} \& \au{Ding, Z.}} \yr{2021}  \at{Coating flows down a
  vertical fibre: towards the full {N}avier--{S}tokes problem}.  \jt{J. Fluid
  Mech.}  \bvol{914},  \pg{A30}.

\bibitem[Maali \& Bhushan(2012)]{maali2012measurement}
{\sc \au{Maali, A.} \& \au{Bhushan, B.}} \yr{2012}  \at{Measurement of slip
  length on superhydrophobic surfaces}.  \jt{Philos. Trans. Royal Soc. A}
  \bvol{370}~(1967),  \pg{2304--2320}.

\bibitem[Maali {\em et~al.\/}(2016)Maali, Colin \& Bhushan]{maali2016slip}
{\sc \au{Maali, A.}, \au{Colin, S.} \& \au{Bhushan, B.}} \yr{2016}  \at{Slip
  length measurement of gas flow}.  \jt{Nanotechnology}  \bvol{27}~(37),
  \pg{374004}.

\bibitem[Mart{\'\i}nez-Calvo {\em et~al.\/}(2020)Mart{\'\i}nez-Calvo,
  Moreno-Boza \& Sevilla]{martinez2020effect}
{\sc \au{Mart{\'\i}nez-Calvo, A.}, \au{Moreno-Boza, D.} \& \au{Sevilla, A.}}
  \yr{2020}  \at{The effect of wall slip on the dewetting of ultrathin films on
  solid substrates: Linear instability and second--order lubrication theory}.
  \jt{Phys. Fluids}  \bvol{32}~(10),  \pg{102107}.

\bibitem[Mostert \& Deike(2020)]{mostert2020inertial}
{\sc \au{Mostert, W.} \& \au{Deike, L.}} \yr{2020}  \at{Inertial energy
  dissipation in shallow--water breaking waves}.  \jt{J. Fluid Mech.}
  \bvol{890},  \pg{A12}.

\bibitem[Plateau(1873)]{plateau1873}
{\sc \au{Plateau, J. A.~F.}} \yr{1873} {\em Statique exp{\'e}rimentale et
  th{\'e}orique des liquides soumis aux seules forces mol{\'e}culaires\/}, ,
  \vol{vol.~2}.  \publ{Gauthier-Villars}.

\bibitem[Popinet(2014)]{popinet2014basilisk}
{\sc \au{Popinet, S.}} \yr{2014}  \at{Basilisk}.  \jt{URl: http://basilisk.
  fr.(accessed: 10.21. 2019)} .

\bibitem[Popinet(2018)]{popinet2018numerical}
{\sc \au{Popinet, S.}} \yr{2018}  \at{Numerical models of surface tension}.
  \jt{Annu. Rev. Fluid Mech.}  \bvol{50},  \pg{49--75}.

\bibitem[Qu{\'e}r{\'e}(1990)]{quere1990thin}
{\sc \au{Qu{\'e}r{\'e}, D.}} \yr{1990}  \at{Thin films flowing on vertical
  fibers}.  \jt{Europhys. Lett.}  \bvol{13}~(8),  \pg{721}.

\bibitem[Qu{\'e}r{\'e}(1999)]{quere1999fluid}
{\sc \au{Qu{\'e}r{\'e}, D.}} \yr{1999}  \at{Fluid coating on a fiber}.
  \jt{Annu. Rev. Fluid Mech.}  \bvol{31},  \pg{347--384}.

\bibitem[Rayleigh(1878)]{rayleigh1878instability}
{\sc \au{Rayleigh, L.}} \yr{1878}  \at{On the instability of jets}.  \jt{Proc.
  London Math. Soc.}  \bvol{1},  \pg{4--13}.

\bibitem[Rayleigh(1892)]{rayleigh1892xvi}
{\sc \au{Rayleigh, L.}} \yr{1892}  \at{Xvi. on the instability of a cylinder of
  viscous liquid under capillary force}.  \jt{London, Edinburgh, Dublin Philos.
  Mag. J. Sci.}  \bvol{34},  \pg{145--154}.

\bibitem[Ruyer-Quil {\em et~al.\/}(2008)Ruyer-Quil, Treveleyan,
  Giorgiutti-Dauphin{\'e}, Duprat \& Kalliadasis]{ruyer2008modelling}
{\sc \au{Ruyer-Quil, C.}, \au{Treveleyan, P.}, \au{Giorgiutti-Dauphin{\'e},
  F.}, \au{Duprat, C.} \& \au{Kalliadasis, S.}} \yr{2008}  \at{Modelling film
  flows down a fibre}.  \jt{J. Fluid Mech.}  \bvol{603},  \pg{431--462}.

\bibitem[Secchi {\em et~al.\/}(2016)Secchi, Marbach, Nigu{\`e}s, Stein, Siria
  \& Bocquet]{secchi2016massive}
{\sc \au{Secchi, E.}, \au{Marbach, S.}, \au{Nigu{\`e}s, A.}, \au{Stein, D.},
  \au{Siria, A.} \& \au{Bocquet, L.}} \yr{2016}  \at{Massive radius--dependent
  flow slippage in carbon nanotubes}.  \jt{Nature}  \bvol{537}~(7619),
  \pg{210--213}.

\bibitem[Tomo {\em et~al.\/}(2022)Tomo, Nag \& Takamatsu]{tomo2022observation}
{\sc \au{Tomo, Y.}, \au{Nag, S.} \& \au{Takamatsu, H.}} \yr{2022}
  \at{Observation of interfacial instability of an ultrathin water film}.
  \jt{Phys. Rev. Lett.}  \bvol{128}~(14),  \pg{144502}.

\bibitem[Tomotika(1935)]{tomotika1935instability}
{\sc \au{Tomotika, S.}} \yr{1935}  \at{On the instability of a cylindrical
  thread of a viscous liquid surrounded by another viscous fluid}.  \jt{Proc.
  R. Soc. Lond. A Math. Phys. Sci.}  \bvol{150}~(870),  \pg{322--337}.

\bibitem[Wei {\em et~al.\/}(2019)Wei, Tsao \& Chu]{wei2019slipping}
{\sc \au{Wei, H.}, \au{Tsao, H.} \& \au{Chu, K.}} \yr{2019}  \at{Slipping
  moving contact lines: critical roles of de gennes’s ‘foot’in dynamic
  wetting}.  \jt{J. Fluid Mech.}  \bvol{873},  \pg{110--150}.

\bibitem[Yu \& Hinch(2013)]{yu2013velocity}
{\sc \au{Yu, L.} \& \au{Hinch, J.}} \yr{2013}  \at{The velocity of
  ‘large’viscous drops falling on a coated vertical fibre}.  \jt{J. Fluid
  Mech.}  \bvol{737},  \pg{232--248}.

\bibitem[Zeng {\em et~al.\/}(2017)Zeng, Sadeghpour, Warrier \&
  Ju]{zeng2017experimental}
{\sc \au{Zeng, Z.}, \au{Sadeghpour, A.}, \au{Warrier, G.} \& \au{Ju, Y.~S.}}
  \yr{2017}  \at{Experimental study of heat transfer between thin liquid films
  flowing down a vertical string in the {R}ayleigh--{P}lateau instability
  regime and a counterflowing gas stream}.  \jt{Int. J. Heat Mass Transf.}
  \bvol{108},  \pg{830--840}.

\bibitem[Zhang {\em et~al.\/}(2022)Zhang, Zheng, Zhu, Zhu, Si \&
  Xu]{zhang2022combinational}
{\sc \au{Zhang, M.}, \au{Zheng, Z.}, \au{Zhu, Y.}, \au{Zhu, Z.}, \au{Si, T.} \&
  \au{Xu, R.~X.}} \yr{2022}  \at{Combinational biomimetic microfibers for
  high-efficiency water collection}.  \jt{Chem. Eng. J.}  \bvol{433},
  \pg{134495}.

\bibitem[Zhang {\em et~al.\/}(2020)Zhang, Sprittles \&
  Lockerby]{zhang2020nanoscale}
{\sc \au{Zhang, Y.}, \au{Sprittles, J.~E.} \& \au{Lockerby, D.~A.}} \yr{2020}
  \at{Nanoscale thin-film flows with thermal fluctuations and slip}.  \jt{Phys.
  Rev. E}  \bvol{102}~(5),  \pg{053105}.

\bibitem[Zhang {\em et~al.\/}(2021)Zhang, Sprittles \&
  Lockerby]{zhang2021thermal}
{\sc \au{Zhang, Y.}, \au{Sprittles, J.~E.} \& \au{Lockerby, D.~A.}} \yr{2021}
  \at{Thermal capillary wave growth and surface roughening of nanoscale liquid
  films}.  \jt{J. Fluid Mech.}  \bvol{915},  \pg{A135}.

\bibitem[Zhao {\em et~al.\/}(2022)Zhao, Liu, Lockerby \&
  Sprittles]{zhao2022fluctuation}
{\sc \au{Zhao, C.}, \au{Liu, J.}, \au{Lockerby, D.~A.} \& \au{Sprittles,
  J.~E.}} \yr{2022}  \at{Fluctuation--driven dynamics in nanoscale thin--film
  flows: physical insights from numerical investigations}.  \jt{Phys. Rev.
  Fluids}  \bvol{7}~(2),  \pg{024203}.

\bibitem[Zhao {\em et~al.\/}(2019)Zhao, Sprittles \&
  Lockerby]{zhao2019revisiting}
{\sc \au{Zhao, C.}, \au{Sprittles, J.~E.} \& \au{Lockerby, D.~A.}} \yr{2019}
  \at{Revisiting the {R}ayleigh--{P}lateau instability for the nanoscale}.
  \jt{J. Fluid Mech.}  \bvol{861}.

\end{thebibliography}

\end{document}